# Surface-induced ordering and continuous breaking of translational symmetry in conjugated polymers

Anton Sinner, Alexander J. Much, Oleksandr Dolynchuk*

*Experimental Polymer Physics, Institute of Physics, Martin Luther University Halle-Wittenberg, Halle D-06120, Germany*

**ABSTRACT**. Surface-induced liquid crystalline phase transitions evoke fundamental interest and can provide deeper insight into the nature of low-dimensional phases. Board-like conjugated polymers are of particular interest because they exhibit novel sanidic liquid crystalline mesophases that have not been widely studied. Furthermore, films of these polymers often exhibit preferential molecular orientation near the free surface, suggesting that the surface influences the ordering process. However, the underlying mechanism of this surface-induced ordering remains unclear. Here, we use surface-sensitive grazing-incidence X-ray scattering to monitor the formation and disappearance of positional order in situ in thin films of two board-like conjugated polymers representative of two classes: polythiophenes and polydiketopyrrolopyrroles. The results show that the free surface induces the formation of a highly oriented, sanidic disordered mesophase at the surface of both conjugated polymers, thus marking the emergence of positional order. This ordering process continues upon cooling and involves multiple transitions into more ordered liquid crystalline mesophases, both at the surface and in the bulk. The positional smectic-like order parameter exhibits continuous temperature dependence near the transition, signifying that the breaking of translational symmetry by the surface occurs continuously. Theoretical analysis allows us to accurately describe the temperature-dependent order parameter when critical behavior is considered.

The existence of long-range positional and orientational order is commonly considered a key criterion for determining the phase state of matter. Liquid crystals (LCs) are an intermediate state between crystals and liquids because they permit orientational and positional order only along specific axes. As a result, they lose 3D translational symmetry and exhibit a variety of structural forms, ranging from nematic to smectic to cholesteric [1]. Due to their inherent anisotropy, the anisotropic mechanical and optical properties of LCs, combined in novel ways, are readily used for a variety of applications [2,3]. Furthermore, the appearance and progression of order in thermotropic LCs can occur through multiple transitions into different phases, elucidating the complex phase diagram of matter and enabling its investigation at a fundamental level [4-6]. Previous studies have shown that geometric and, more broadly, topological constraints are crucial to the nature of phase transitions and the behavior of the order parameter [7-14]. In particular, surfaces and interfaces naturally impose positional constraints on the system. These constraints affect positional and orientational order, leading to phase transitions such as surface melting, layer thinning, surface freezing, prefreezing, and prewetting [15-26]. Moreover, the influence of surfaces can be diverse. As demonstrated by Lipowsky et al., surface freezing is a first-order phase transition, whereas surface melting is a second-order phase transition [14].

Surface-induced phase transitions have been observed in various systems and materials, including inorganic and organic crystals, ionic liquids, and different types of thermotropic LCs [15,20,27-30]. While surface melting is commonly found in inorganic solids, surface freezing is more prevalent in organic LCs [15,16]. For example, rod-shaped organic mesogens can form freely suspended films with surface-stabilized smectic phases [19]. The smectic layers in these films align parallel to the free surface, making surface-induced order experimentally accessible using X-ray scattering. Specifically, the in-plane order of different orthogonal smectic phases, ranging from liquid to hexatic to crystalline, as well as the associated phase transitions in the bulk and at the surface can be studied by varying the temperature and film thickness [19]. Furthermore, very thin smectic films approach two-dimensional behavior, making them ideal model systems for studying the differences between two- and three-dimensional systems. Nevertheless, providing a comprehensive theoretical description of these surface-induced phase transitions, including low-dimensional ones, remains challenging, especially regarding the order of the phase transition [19]. In this context, a new class of LCs called sanidic LCs, which are found in board-like polymers consisting of a stiff backbone with short side chains attached (see Fig. 1(a)), is of particular interest [31]. Similar to smectic LC mesophases, sanidic LCs exhibit orientational and positional order in specific directions and can form layers. However, the covalent bonds between monomers inside the polymer chain introduce additional in-plane constraints and correlations in nearest-neighbor interactions. This leads to additional in-plane phases not observable in smectic LCs [31]. The backbone-side-chain architecture is also common among modern conjugated polymers, such as poly(3-hexylthiophene) (P3HT) and copolymers of diketopyrrolopyrroles (PDPPs), which also belong to the board-like class and have been shown to exhibit a variety of sanidic LC mesophases [32, 33]. Notably, board-like conjugated polymers are of particular interest in the context of surface-induced ordering, as they often exhibit a preferred out-of-plane crystal orientation in thin films, suggesting that surfaces influence their structure formation. The first studies of surface-induced ordering in P3HT films indicated that the top free surface induces an edge-on molecular orientation, with alkyl side chains perpendicular to the film surface, independent of the substrate [34,35]. This pattern resembles the surface freezing of short alkanes and alkyl side-chain polymers [20,28-30]. Building on these findings, recent work has examined the crystallization of P3HT films in situ, concluding that crystal nucleation is the ordering mechanism at the free surface [36]. However, no LC mesophases were observed, conflicting with earlier observations in P3HT bulk [32].

*Contact author: oleksandr.dolynchuk@physik.uni-halle.de

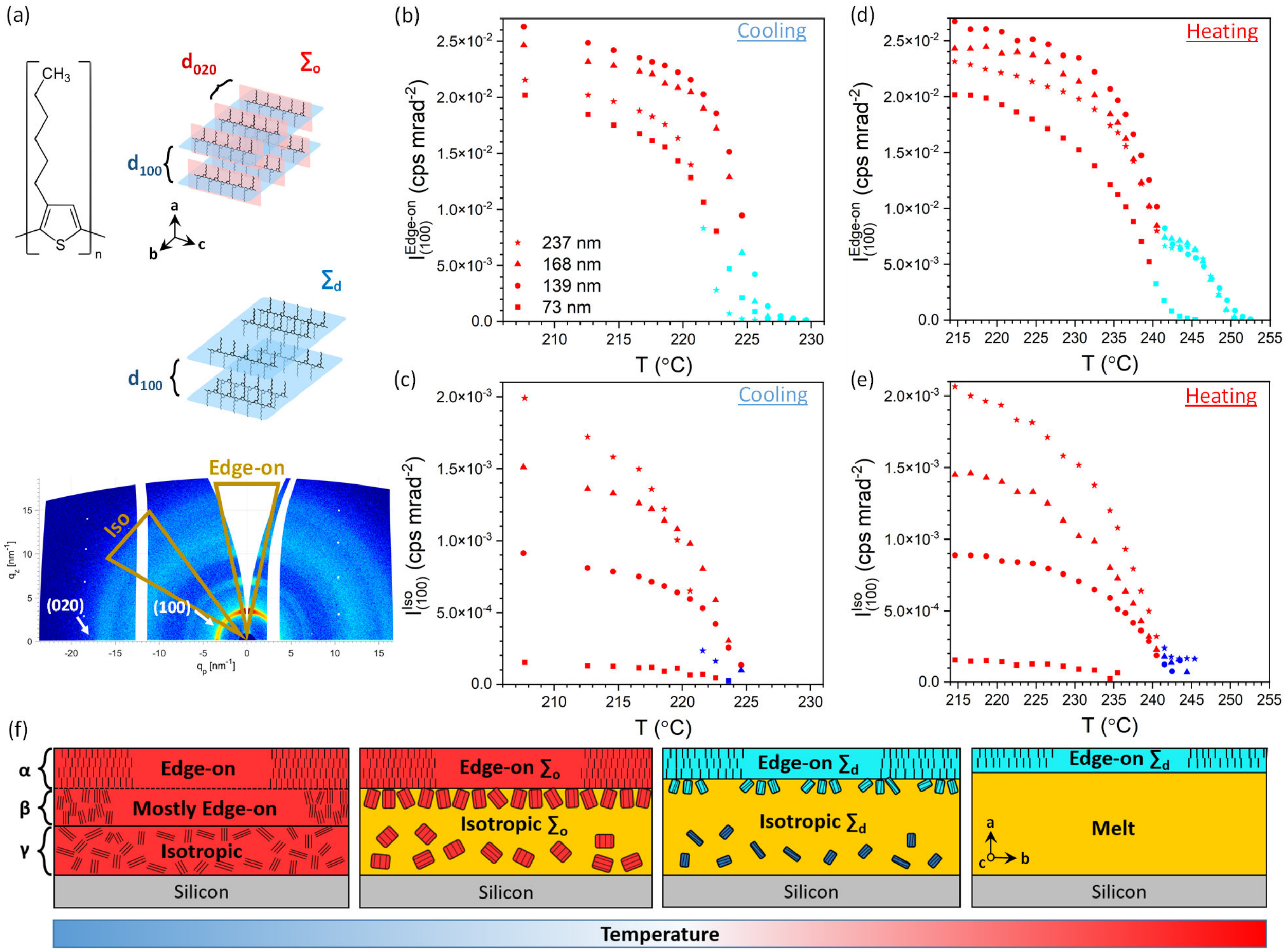

FIG. 1. (a) Chemical structure of P3HT along with a sketch of the $\Sigma_o$ and $\Sigma_d$ LC mesophases in P3HT and an exemplary GIWAXS image of P3HT film on silicon with $L = 237$ nm. The $\Sigma_o$ LC mesophase signifies positional order along the side-chain direction (blue planes) and in-between chains in the π-π stacking direction (red planes). The positional order in the π-π stacking direction is lost in the $\Sigma_d$ LC mesophase. (b-e) The intensity $I_{(100)}$ vs. temperature of edge-on oriented (b,d) and isotropic (c,e) LCs measured in P3HT films on silicon during cooling (b,c) and heating (d,e). The different symbols indicate different film thicknesses, as given in (b). Blue and cyan indicate the bulk and surface $\Sigma_d$ LC mesophases, respectively, and red indicates the $\Sigma_o$ LC mesophase. (f) A sketch showing the ordering of P3HT films. The (00l) plane is shown (compare to (a)). The short lines inside the α, β and γ regions represent side chains. The polymer backbones are oriented perpendicular to the sketch plane.

In light of all prior observations, the following questions arise naturally: How does the free surface of the film induce positional order in board-like conjugated polymers, particularly P3HT and the representative PDPP, PDPP[T]$_2$-T? Are there observable thermotropic LC mesophases in thin films, and how does the surface affect their appearance? Furthermore, it is essential to determine whether surface-induced phase transitions in board-like conjugated polymers can be explained by existing phenomenological Landau-type theories. In this study, we use grazing-incidence wide-angle X-ray scattering (GIWAXS) at high temperatures to monitor the formation and equilibrium state of the structure of board-like conjugated polymer films in situ. This method allows us to track the positional order in polymer films in different spatial directions, thereby isolating the surface influence and directly addressing the above questions.

*Contact author: oleksandr.dolynchuk@physik.uni-halle.de

First, the ordering in films of the model board-like conjugated polymer P3HT was investigated (see Fig. 1(a)). Thin films of P3HT with a molecular weight of $M_n$=15.6 kg/mol were prepared on a silicon substrate by spin-coating from chloroform solutions. The film thickness, $L$, was varied, ranging from 73 to 237 nm (see Supplemental Material (SM) for details) [37]. After heating and melting the films, the ordering was monitored during slow cooling from the melt with long quasi-isothermal steps. At each isothermal step, the samples were equilibrated for 30 minutes and then measured using GIWAXS for an additional 30 minutes (see SM for details) [37]. The resulting GIWAXS patterns were analyzed in different spatial directions (see Fig. 1(a) and SM for the analysis details) to isolate the influence of the surface from that of the bulk [37]. Specifically, we quantified the temperature evolution of the integrated intensities $I_{(100)}^{Edge-on}$ and $I_{(100)}^{Iso}$, which are scattered from edge-on oriented and isotropic (100) layers of polymer chains with a periodic distance $d_{(100)}$ at the surface and in the bulk, respectively. The intensity $I_{(100)}$ is directly proportional to the number of (100) layers and the squared form factor of each layer: $I_{(100)}(T) \sim N_{(100)}(T) \cdot |F_{(100)}(T)|^2$, where $N_{(100)}(T)$ and $|F_{(100)}(T)|$ are both temperature dependent. Additionally, the presence of other peaks in the GIWAXS pattern was monitored, particularly the (020) peak, which represents π-π stacking between the backbones within the (100) layers, to determine the phase state at different temperatures. If only the (100) peak is present, it indicates positional order only along the side-chain direction and a sanidic disordered $\Sigma_d$ LC mesophase. The appearance of the (020) peak marks the formation of additional positional order between the polymer backbones within the (100) layers and a sanidic ordered $\Sigma_o$ LC mesophase (see Fig. 1(a)). Fig. 1(b, c) shows the temperature evolution of the intensities $I_{(100)}^{Edge-on}$ and $I_{(100)}^{Iso}$ during cooling from the melt for different P3HT film thicknesses, which are represented by different symbols. The edge-on and bulk $\Sigma_d$ LC mesophases are distinguished by cyan and blue colors, respectively, whereas the edge-on and bulk $\Sigma_o$ LC mesophases are indicated by red. Fig. 1(b) shows that the formation of the edge-on $\Sigma_d$ LC mesophase at 230-228°C marks the onset of positional order, regardless of film thickness. This phase grows slowly with decreasing temperature until a steep increase occurs at 225-223°C. Within this temperature range, the isotropic bulk $\Sigma_d$ LC mesophase forms as well (Fig. 1(c)). While the appearance of $I_{(100)}^{Edge-on}$ is continuous, $I_{(100)}^{Iso}$ features a jump that is more visible in thicker films. A few degrees below, both the edge-on and isotropic bulk $\Sigma_d$ mesophases undergo transitions into their respective $\Sigma_o$ LC mesophases and continue to grow rapidly. Apparently, the two ordering processes compete with each other at this stage. Additional surface-sensitive GIWAXS measurements performed in this temperature range and at room temperature confirm the simultaneous growth of oriented and bulk LCs and reveal that the edge-on LC mesophase dominates near the free surface, as expected [37]. Furthermore, as shown in Fig. 1, $I_{(100)}^{Edge-on}$ and $I_{(100)}^{Iso}$ exhibit different dependencies on film thickness after the ordering process. Despite a 300% variation in film thickness, the intensity $I_{(100)}^{Edge-on}$ changes by only about 15% between films, primarily due to scattering from the film edges [37]. In contrast, the intensity $I_{(100)}^{Iso}$ increases by about one order of magnitude with increasing film thickness $L$. Together, these results suggest that the oriented crystals have a constant layer thickness at room temperature, confirming earlier results [34]. Nonetheless, our findings reveal that the constancy of the edge-on-oriented layer stems from the competition between surface-induced, edge-on-oriented, and bulk, isotropic, ordering processes that occur through multiple sanidic LC mesophases. Note that the transition from the $\Sigma_o$ mesophase to the crystal phase at lower temperatures in Fig. 1 was not visible due to the low intensity of the representative $(1\bar{1}1)$ peak for this range of film thicknesses. However, the $(1\bar{1}1)$ peak can be identified in the GIWAXS patterns of a 237 nm thick P3HT film at room temperature, and the respective transition does occur in the bulk [37].

To better understand the resulting vertical film morphology and study the equilibrium state of the structure, the melting of the same P3HT films was monitored by in situ GIWAXS during stepwise heating, similar to the cooling process. Fig. 1(d) illustrates the temperature dependence of $I_{(100)}^{Edge-on}$ during heating. All films with a thickness of $L \geq 139$ nm demonstrate an identical two-step melting profile. In the temperature range of 235-241°C, the intensity

*Contact author: oleksandr.dolynchuk@physik.uni-halle.de

$I_{(100)}^{Edge-on}$ drops by roughly 70-80%, accompanied by a decrease in the azimuthal distribution of $I_{(100)}^{Edge-on}$, thus only leaving highly oriented edge-on $\Sigma_o$ LCs [37]. In the same temperature range, as seen in Fig. 1(e), the intensity $I_{(100)}^{Iso}$ of the bulk $\Sigma_o$ LCs behaves similarly, indicating that the bulk and large amounts of edge-on oriented LCs have the same thermal stability. At higher temperatures, both edge-on and bulk $\Sigma_o$ LCs undergo transitions into their respective $\Sigma_d$ phases. The bulk $\Sigma_d$ mesophase abruptly melts at around 247°C, indicating a weak first-order phase transition, whereas the surface-induced $\Sigma_d$ mesophase melts continuously up to 253-254°C. Remarkably, the results from different films overlap well in this region, clearly indicating the independence of the surface influence on film thickness for $L \geq 139$ nm. Together, these findings demonstrate that the free surface of a film enhances the thermal stability of the edge-on-oriented $\Sigma_d$ LC mesophase and modifies the phase transition. Furthermore, the two-step nature of $I_{(100)}^{Edge-on}$ during melting, as opposed to its behavior during cooling, suggests that two edge-on oriented layers formed: one with a high melting point and the other with bulk thermal stability, referred to as the α- and β-layers, respectively. A comparison of the results of cooling and heating experiments on thick P3HT films in Fig. 1(b)-(e) leads to the ordering model proposed in Fig. 1(f). During cooling, the free surface induces the formation of an edge-on oriented $\Sigma_d$ mesophase, or α-layer. After several hours and multiple temperature steps, the bulk below the growing α-layer begins to order into the isotropic $\Sigma_d$ mesophase, forming the γ-layer. At the same time, the edge-on-oriented $\Sigma_d$ mesophase begins to grow steeply, indicating the formation of a bulk-like, edge-on-oriented β-layer at the front of the α-layer – similar to a self-nucleation process. The β-layer has a loose orientation, unlike the α-layer, making it mostly edge-on oriented [37]. While growing. the β-layer may hinder the growth of the α-layer and compete with the γ-layer. The existence of the β-layer and its relative thickness to the α-layer are clearly seen during subsequent heating, when $I_{(100)}^{Edge-on}$ exhibits a two-step behavior. However, for thinner films $L \leq 73$ nm, as shown in Fig. 1(b)-1(e) and the SM, the ordering during cooling and the melting during heating occur in one step, though they show the same sequence of sanidic LC mesophases as their thicker counterparts [37]. Additionally, the temperatures of the LC phase transitions at the free surface and in the bulk gradually decrease with decreasing $L$, likely indicating confinement effects or the influence of the silicon interface, which comes closer to the film surface as $L$ decreases. Furthermore, the results from thinner, fully edge-on oriented films were used to estimate the thickness of the α-layer in the $\Sigma_d$ state after the γ-layer melted, which amounts to approximately 12 nm, or 6 polymer layers [37]. Finally, a comparison of the results from the cooling and heating experiments in Fig. 1 clearly shows a significant temperature hysteresis of about 20-25°C between the formation and melting of the respective $\Sigma_d$ phases at the surface and in the bulk. Generally, temperature hysteresis is not surprising for the formation of the bulk $\Sigma_d$ phase, as it is a first-order phase transition that requires finite supercooling to overcome an energy barrier. However, the clear, continuous nature of $I_{(100)}^{Edge-on}$ at high temperatures rules out a first-order transition. This is difficult to reconcile with the significant supercooling required for the formation of the edge-oriented $\Sigma_d$ phase. Although the P3HT films were kept in a molten state for about six hours during cooling, and the unconstrained growth of the α-layer took at least five hours, we question whether an equilibrium state of the structure was reached.

Thus, to address the presence of the temperature hysteresis and gain deeper insight into the formation of the edge-on $\Sigma_d$ phase, isothermal in situ GIWAXS experiments were performed, the results of which are shown in Fig. 2(a). Furthermore, we tested whether the observed ordering phenomena in P3HT films are independent of molecular weight and truly inherent to P3HT. For this purpose, analogous in situ GIWAXS experiments were performed on thin films of P3HT with $M_n$=47.2 kg/mol, which securely exceeds the entanglement molecular weight of approximately 14 kg/mol [38]. Both P3HT samples exhibit the same sanidic LC mesophases in bulk upon heating, though the higher $M_n$ sample has lower transition temperatures [37]. Fig. 2(a) shows the kinetics of edge-on $\Sigma_d$ mesophase formation (α-layer) measured at high temperatures prior to the formation of the bulk $\Sigma_d$ mesophase (γ-layer). For both P3HT alike, the kinetics slow down significantly after about two hours of initial rapid progression. Notably, the increase in supercooling of the low $M_n$ sample does not change the growth kinetics qualitatively, rather it increases proportionally. Although the ordering process takes over 15 hours, it is still insufficient for the edge-on $\Sigma_d$ phase to grow into the bulk. This growth occurs only during subsequent cooling, which is accompanied by the formation of the mostly edge-on oriented and bulk $\Sigma_d$ phase (β-layer and γ-layer). Additionally, the temperature of the formation of the edge-on $\Sigma_d$ phase is lower for the higher $M_n$ sample. This is consistent with its lower crystallization temperature in the bulk and is apparently due to the slower dynamics of entangled chains [37]. To observe the effects of isothermal ordering on the resulting structure, the samples were heated, and their melting behavior was compared to that after stepwise cooling. Contrary to cooling, introducing isothermal measurement steps during heating shows no time dependence of melting, thereby indicating an equilibrium state [37]. As shown in Fig. 2(b), the results clearly demonstrate the formation of a thicker edge-on oriented layer (α-layer and β-layer) and a thicker α-layer inside it after isothermal ordering, which becomes dominant for the low $M_n$ sample. Here, the melting temperatures of the edge-on $\Sigma_o$ and $\Sigma_d$ mesophases remain nearly unaffected by the cooling protocol. However, they are higher after isothermal ordering in the high-$M_n$ sample and close to the melting temperature of the low-$M_n$ P3HT. This appears to reflect the complex influence of slower chain dynamics on the ordering in this sample. Together, the results in Fig. 2(a),(b) show that isothermal ordering enhances the growth of the edge-on $\Sigma_d$ mesophase (α-layer) and affects its competition with the bulk $\Sigma_d$ mesophase (γ-layer), resulting in a thicker edge-on oriented layer. Furthermore, surface-induced ordering is qualitatively independent of the $M_n$ in P3HT and differs only quantitatively. Using a monotonic dependence of the intensity $I_{(100)}^{Iso}$ on $L$ at room temperature, we determined the thickness of the edge-oriented layer after continuous cooling to be approximately 60 and 40 nm for P3HT with $M_n$=15.6 kg/mol and $M_n$=47.2 kg/mol, respectively [37]. Finally, an examination of the surface morphology of both P3HT samples with AFM and POM revealed long, extended lamellae on the surface without any nucleation sites, further supporting the preferential vertical growth of the edge-on $\Sigma_d$ mesophase (α-layer) during cooling [37].

*Contact author: oleksandr.dolynchuk@physik.uni-halle.de

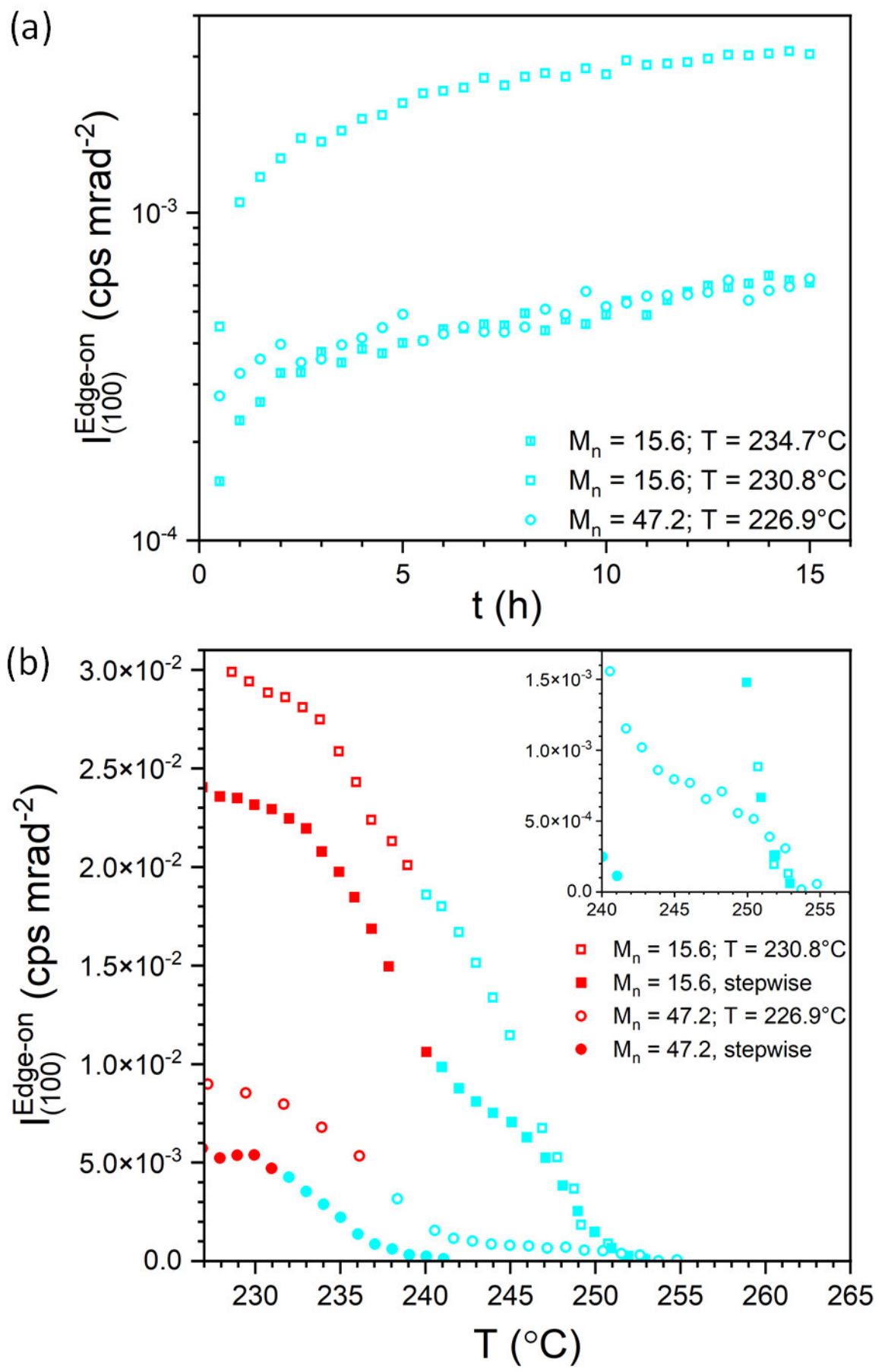

FIG 2. (a) Time evolution of the intensity $I_{(100)}^{Edge-on}$ during isothermal ordering of P3HT films on silicon ($L$≈130 nm) with $M_n$=15.6 kg/mol and $M_n$=47.2 kg/mol. Measurements were performed isothermally after cooling from the melt at temperatures indicated in the legend. (b) $I_{(100)}^{Edge-on}$ vs. temperature during heating of P3HT films on silicon ($L$≈130 nm) with $M_n$=15.6 kg/mol and $M_n$=47.2 kg/mol. The films were ordered either isothermally, as shown in (a), or after stepwise cooling. The inset enlarges the melting of the α-region. Red and cyan indicate the $\Sigma_o$ and $\Sigma_d$ LC states, respectively.

To see how general the surface-induced ordering observed in P3HT films is, we turn our attention to another board-like conjugated polymer: PDPP[T]$_2$-T (Fig. 3(a)) [33]. Unlike P3HT, PDPP[T]$_2$-T is only liquid-crystalline at room temperature, specifically it forms the $\Sigma_o$ mesophase. When ordered from the melt, the polymer shows only edge-on orientation in 100 nm thick films on silicon, indicating an even stronger influence of the surface [33]. Therefore, similar to the above studies on P3HT films, in situ GIWAXS measurements were performed on thicker PDPP[T]$_2$-T films on silicon with $L$=286 and 469 nm during cooling and heating to allow the surface influence to extend. Throughout the entire high-temperature range of the experiment, only the scattering from edge-on oriented LCs could be observed, and the intensity $I_{(100)}^{Edge-on}$ is plotted versus temperature in Fig. 3(b). Similarly to P3HT, positional order in PDPP[T]$_2$-T films begins with the formation of the edge-on $\Sigma_d$ mesophase at the film surface. Unlike P3HT, the formation of the edge-on $\Sigma_d$ mesophase in PDPP[T]$_2$-T occurs 8 to 20°C above the bulk melting point [33]. At lower temperatures and after further growth, this phase undergoes a transition into the $\Sigma_o$ mesophase. During successive melting, the reverse order of phase transitions occurs, with a temperature hysteresis of about 10°C for the $\Sigma_o$ mesophase and a notably larger 16-27°C for the $\Sigma_d$ mesophase. Remarkably, the equilibrium temperature range of the $\Sigma_d$ mesophase during heating extends by 22°C in the thinner film and by 49°C in the thicker film, unequivocally indicating continuous melting of the positional order. The single sigmoidal shape of $I_{(100)}^{Edge-on}$ during heating suggests that the entire edge-on oriented layer was formed by the growth of the surface-induced $\Sigma_d$ phase (α-layer) during cooling. Room-temperature GIWAXS revealed a small signal from the bulk $I_{(100)}^{Iso}$ (γ-layer) (see SM), which formed and melted during cooling and heating below the temperature range shown in Fig. 3(b) [37]. This signal was used to estimate the thickness of the edge-on oriented layer in PDPP[T]$_2$-T to be about 180 nm – three times thicker than in P3HT [37]. Overall, these results demonstrate the same multistep, surface-induced LC ordering in PDPP[T]$_2$-T as in P3HT, but with a notably greater surface influence, thus revealing the generality of the phenomenon for board-like conjugated polymers.

*Contact author: oleksandr.dolynchuk@physik.uni-halle.de

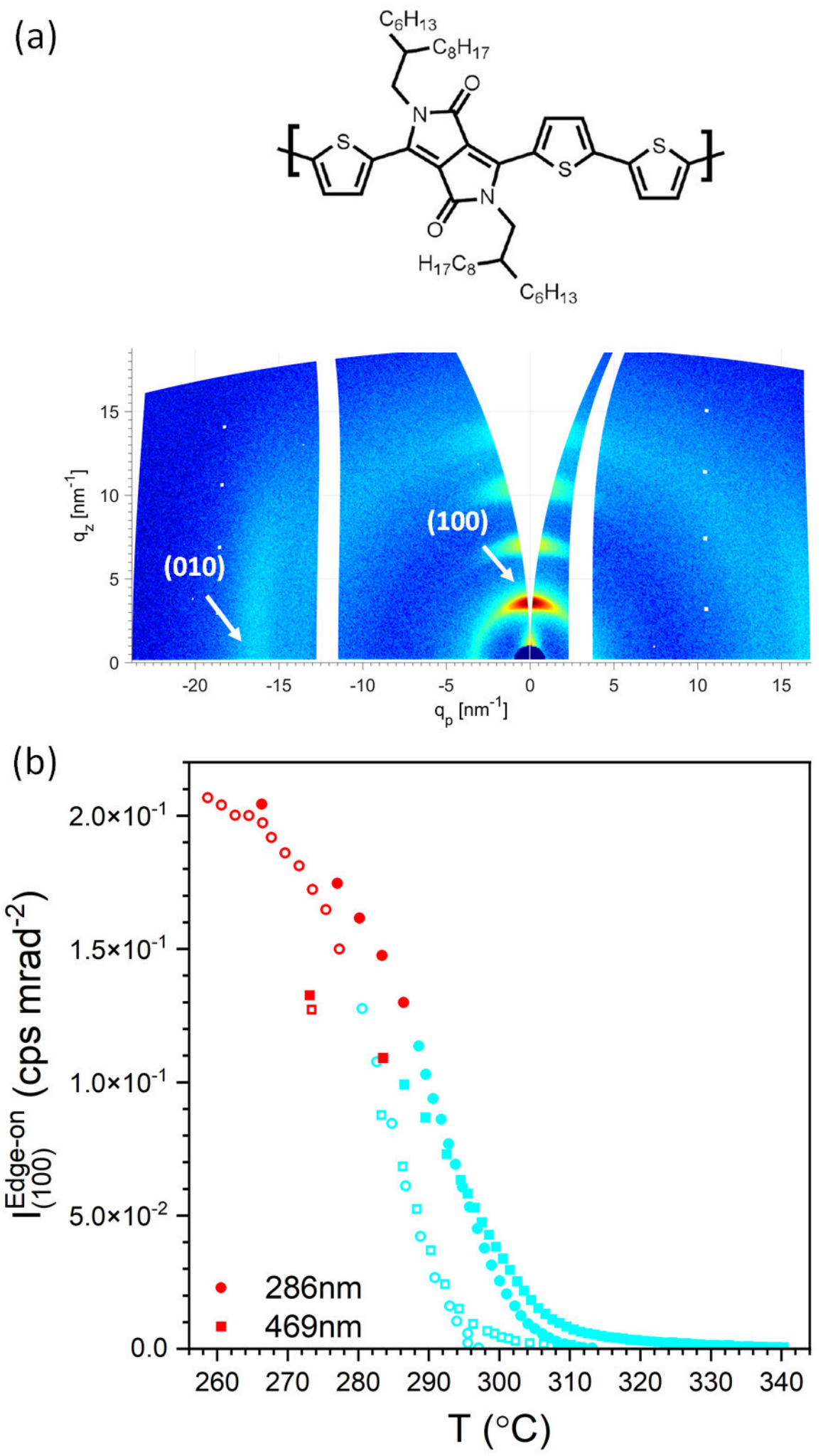

FIG 3. (a) Chemical structure of PDPP[T]$_2$-T and exemplary GIWAXS image of PDPP[T]$_2$-T film on silicon with $L$ = 469 nm. (b) Intensity $I_{(100)}^{Edge-on}$ of edge-on oriented LCs vs. temperature measured in PDPP[T]$_2$-T films on silicon with $L$ = 286 (circles) and 469 nm (squares) during cooling (empty symbols) and heating (filled symbols). Red and cyan indicate a $\Sigma_o$ and $\Sigma_d$ LC state, respectively.

We now wish to gain more clarity about how the surface affects the appearance of positional order in board-like conjugated polymers and the corresponding phase transition type. As discussed, the $I_{(100)}$ intensities presented thus far depend on the number of (100) planes, or polymer layers (Fig. 1(a)), and the form factor of the (100) planes, which represents the average order within the layers. This type of positional order is equivalent to that in smectic LCs and can be described by the smectic order parameter $|\Psi|$, defined as the amplitude of the density wave: $\Psi = |\Psi| e^{ik_0 z}$, $k_0 = \frac{2\pi}{d_{(100)}}$. As is often done in studies of smectic LCs, the intensity scattered from the layered structure, normalized by its largest value at low temperatures, yields the square of the smectic order parameter [11,39,40]:

$$\frac{I_{(100)}(T)}{I_{(100)}^{max}} = \frac{N_{(100)}(T)\cdot\left|F_{(100)}(T)\right|^2}{N_{(100)}^{max}\cdot\left|F_{(100)}^{max}\right|^2} \approx |\Psi|^2. \qquad (1)$$

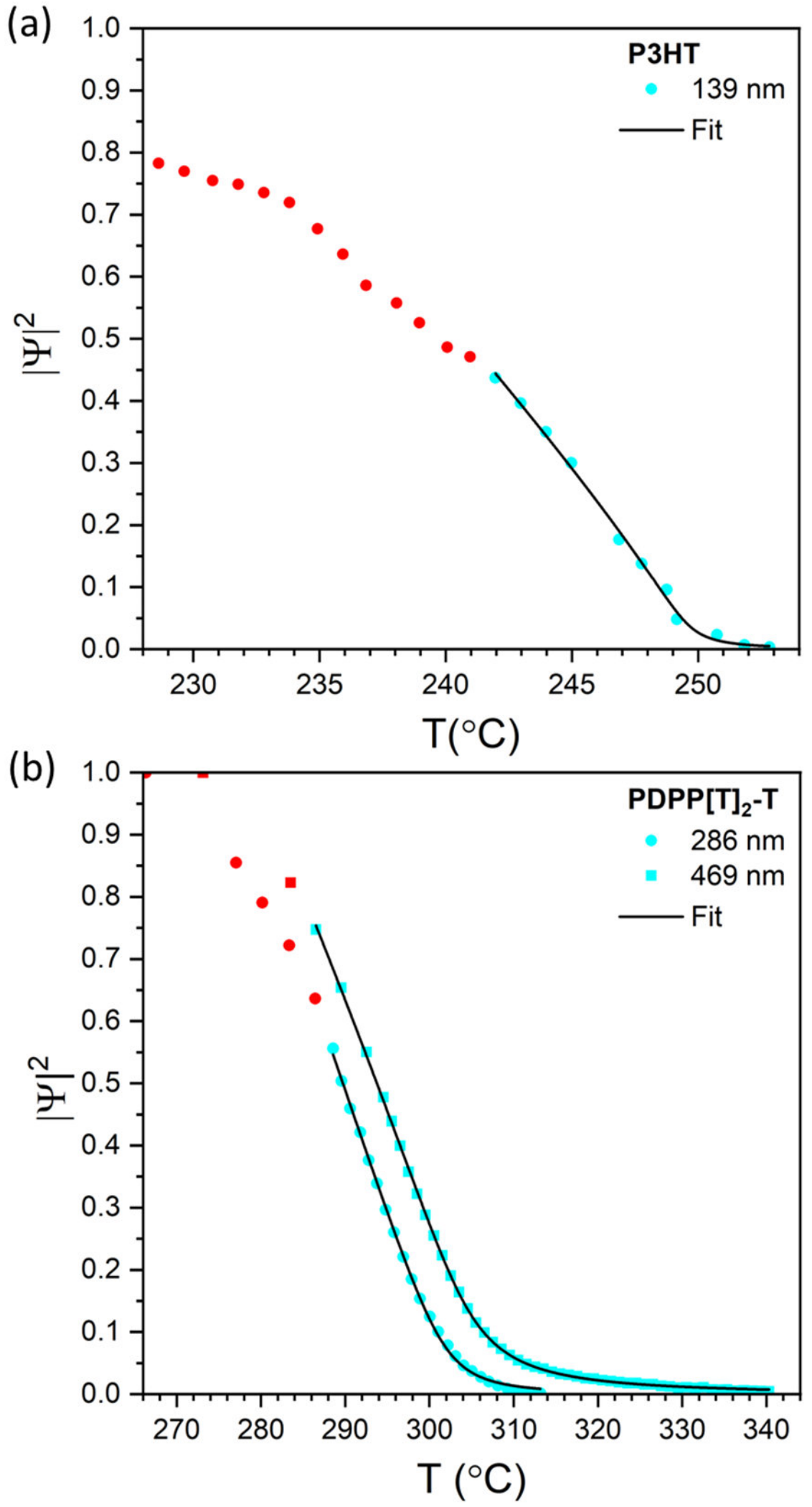

FIG 4. The temperature dependence of the smectic order parameter, $|\Psi|^2$, during the melting of edge-on oriented LCs in P3HT (a) and PDPP[T]$_2$-T (b) films on silicon, as obtained by equation (1). Film thicknesses are given in the figure legends. Red and cyan indicate a $\Sigma_o$ and $\Sigma_d$ LC state, respectively. The cyan data were fitted with the solution to equation (3) (black lines) [37].

*Contact author: oleksandr.dolynchuk@physik.uni-halle.de

Thus, the intensity $I_{(100)}^{Edge-on}$, which was measured during the heating of the P3HT and PDPP[T]$_2$-T films and represents the equilibrium structure, was converted into the positional order parameter $|\Psi|^2$ in Fig. 4. The temperature dependence of $|\Psi|^2$ in the $\Sigma_d$ state (cyan data in Fig. 4) is clearly similar for both polymers. Furthermore, despite a temperature shift, identical behavior of $|\Psi|^2$ is evident for the two PDPP[T]$_2$-T films with different thicknesses. The obtained generic form of $|\Psi|^2$ allows us to apply existing Landau-type theories to describe it. The phenomenological theory of surface-induced freezing was first developed by Lipowsky et al., and then by Frenkel et al., in analogy to the semi-infinite Ising model [14,16,41]. By considering a nonuniform order parameter profile across a semi-infinite film, the temperature- and depth-dependent order parameter profiles can be derived. As our calculations based on this theory illustrate, positional order forms at the surface and propagates into the bulk as the temperature decreases [37]. However, the first-order surface freezing proposed by this theory fails to adequately describe the continuous character of the order parameter in Fig. 4 [37]. The predicted order parameter exhibits a concave shape with a discontinuity at the transition point. Notably, the results in Fig. 4 closely resemble the behavior of the orientational order parameter of nematic LCs in an external field [42,43]. Given that the positional order parameter in Fig. 4 is coupled with the orientational order parameter, it is reasonable to assume that the free surface induces a predominant edge-on molecular orientation prior to the formation of positional order. This would imply a surface-induced nematic phase. Previous studies indeed suggested the existence of nematic phase in the bulk of board-like conjugated polymers [44]. Additionally, nematic wetting of the free surface is readily observed in calamitic organic mesogens and is also predicted by theory [45,46]. Thus, the existence of a nematic phase at the surface of board-like polymers would then result in an effective local molecular field due to the molecular anisotropy of the monomer unit, which in our case bears a permanent dipole [47,48]. The effective molecular field, which is tied to the orientational order parameter, then triggers the formation of positional order. Thus, the analogy between the nematic LCs in an external field and the results in Fig. 4 becomes apparent. The molecular field breaks director symmetry and can be considered linear in $|\Psi|$. Similarly to the description of nematic LCs in an external field [42, 43], adding a negative linear term to the standard expression for the free energy density that describes weak first-order phase transitions yields:

$$f(|\Psi|, T) = f_0 - h|\Psi| + \frac{a}{2}|\Psi|^2 - \frac{b}{3}|\Psi|^3 + \frac{c}{4}|\Psi|^4. \quad (2)$$

By minimizing the expression (2) with respect to $|\Psi|$ and factoring by $c$ to avoid overparametrization of the solution, we obtain:

$$-\tilde{h} + \tilde{a}|\Psi| - \tilde{b}|\Psi|^2 + |\Psi|^3 = 0, \qquad (3)$$

with $\tilde{a} = \frac{a}{c} = \frac{a_0}{c}(T - T_0) = \tilde{a}_0(T - T_0)$, $\tilde{b} = \frac{b}{c}$, $\tilde{h} = \frac{h}{c}$, where $\tilde{h}$ describes the mean molecular field strength. The solution to the cubic equation (3) yields the equilibrium order parameter $|\Psi| = |\Psi|(T)$ [37]. By using this solution to fit our experimental data we obtained an excellent match, as shown in Fig. 4. Furthermore, the fitting parameters listed in SM reveal a four- to sevenfold increase in $\tilde{h}$ for PDPP[T]$_2$-T relative to P3HT [37]. This clearly indicates a stronger surface influence and is consistent with our experimental observations. Additionally, $\tilde{b}$, the parameter responsible for the weak first-order phase transition in the bulk when the linear term in equation (2) is omitted, takes sensible values, which are again consistent with our experimental data [37]. Importantly, the theoretical curve of $|\Psi|^2$ describes a continuous change in the positional order parameter across the entire temperature range of applicability. Similar critical behavior of the positional and orientational order parameter was found for the smectic-isotropic transition under confinement and predicted for the nematic-isotropic transition at surfaces [11,46,49]. Moreover, since ordering begins at the surface, as evidenced by our experimental data and the order parameter profile [14,37,41,46,49], it is reasonable to infer that the continuous decrease in positional order in board-like conjugated polymers during heating occurs down to a single two-dimensional edge-on oriented molecular layer at the surface, reminiscent of layer thinning in free-standing smectic membranes [17-19]. However, not all rotational isomeric states are compatible with an edge-on orientation, thus the polymer chains will unavoidably straighten [50,51]. Due to the monomer connectivity, this will lead to at least quasi long-range orientational and positional order between the monomer units in the polymer chain. Consequently, this will lead to an even higher chain rigidity, especially for PDPP[T]$_2$-T, rendering the semi-flexible chains quasi rod-like [52]. Thus, the polymer chains within this surface layer will necessarily be aligned parallel to each other, at least locally, forming a two-dimensional nematic phase with in-plane orientational order between the polymer chains. Assuming no substantial density changes in this surface layer, in-plane short-range orientational order between the chains is rather unlikely implying long-range or quasi-long range orientational order. According to the Hohenberg-Mermin-Wagner theorem, there can be no associated spontaneous symmetry breaking for such long-range orientational order [53]. However, quasi-long-range order is allowed, implying an associated Berezinskii-Kosterlitz-Thouless (BKT) transition into a topological nematic state [7]. Notably, a BKT transition from a liquid into a two-dimensional nematic phase has been observed in simulations of hard rod fluids [8-10]. Additionally, theoretical considerations of biaxial nematics in close proximity to a wall with a strong surface field lead to the conjecture of a BKT transition at the surface [46]. These findings are highly relevant to our system due to the similarity between hard rods and stiff board-like chains. Therefore, these arguments lead us to hypothesize a topological BKT transition at the surface of board-like conjugated polymers. This transition results in the formation of a single edge-on oriented molecular layer, which then triggers the formation of out-of-plane positional order.

*Contact author: oleksandr.dolynchuk@physik.uni-halle.de

In summary, our results demonstrate that surface-induced ordering is a general phenomenon in board-like conjugated polymer films. This ordering begins at the surface and involves multiple transitions into different sanidic LC mesophases, arising from the board-like architecture of the polymer. The surface-induced positional order obtained from the experimental data exhibits clear, continuous temperature dependence and cannot be described by existing Landau-type theories that assume a first-order phase transition. A distinct temperature hysteresis between the formation and disappearance of positional order excludes a second-order phase transition and allows us to hypothesize a topological BKT-type transition induced by the surface.

## ACKNOWLEDGMENTS

We acknowledge financial support from the Deutsche Forschungsgemeinschaft (DFG, German Research Foundation) – Projektnummer 513509216 and the Ministry of Science, Energy, Climate Protection and Environment of the State of Saxony-Anhalt (grant no. 41-04032/2018). The authors thank Robert T. Kahl for providing the PDPP[T]$_2$-T polymer. We are also grateful to Viktor Ivanov, Thomas Thurn-Albrecht, and Jens-Uwe Sommer for insightful discussions about the results and their critical review of the manuscript.

[1] P. G. De Gennes, The Physics of Liquid Crystals (Clarendon, Oxford, 1974).
[2] I. W. Hamley, Introduction to Soft Matter: Synthetic and Biological Self-Assembling Materials (Wiley, Hoboken, N.J, 2013).
[3] S. Chandrasekhar, Liquid Crystals S. Chandrasekhar (Cambridge Univ. Press, Cambridge, 2003).
[4] P. Biscari, M. C. Calderer, and E. M. Terentjev, Physical Review E **75**, (2007).
[5] P. K. Mukherjee, Journal of Molecular Liquids **190**, 99 (2014).
[6] K. C. Chu and W. L. McMillan, Physical Review A **15**, 1181 (1977).
[7] J. M. Kosterlitz, Reports on Progress in Physics **79**, 026001 (2016).
[8] M. Dijkstra and D. Frenkel, Physical Review E **50**, 349 (1994).
[9] M. A. Bates and D. Frenkel, The Journal of Chemical Physics **112**, 10034 (2000)
[10] D. Frenkel and R. Eppenga, Physical Review A **31**, 1776 (1985).
[11] N. A. Clark et al., Physical Review Letters **71**, 3505 (1993).
[12] P. K. Mukherjee and S. Biswas, Journal of Molecular Liquids **271**, 182 (2018).
[13] M. Heni and H. Löwen, Physical Review Letters **85**, 3668 (2000).
[14] R. Lipowsky and W. Speth, Physical Review B **28**, 3983 (1983).
[15] J. G. Dash, Contemporary Physics **30**, 89 (1989).
[16] B. Pluis, D. Frenkel, and J. F. van der Veen, Surface Science **239**, 282 (1990).
[17] T. Stoebe, P. Mach, and C. C. Huang, Physical Review Letters **73**, 1384 (1994).
[18] P. M. Johnson et al., Physical Review E **55**, 4386 (1997).
[19] W. H. de Jeu, B. I. Ostrovskii, and A. N. Shalaginov, Reviews of Modern Physics **75**, 181 (2003).
[20] B. M. Ocko et al., Physical Review E **55**, 3164 (1997).
[21] O. Gang, X. Z. Wu, B. M. Ocko, E. B. Sirota, and M. Deutsch, Physical Review E **58**, 6086 (1998).

*Contact author: oleksandr.dolynchuk@physik.uni-halle.de

[22] B. M. Ocko, A. Braslau, P. S. Pershan, J. Als-Nielsen, and M. Deutsch, Physical Review Letters **57**, 94 (1986).
[23] A.-K. Löhmann, T. Henze, and T. Thurn-Albrecht, Proceedings of the National Academy of Sciences **111**, 17368 (2014).
[24] O. Dolynchuk, M. Tariq, and T. Thurn-Albrecht, The Journal of Physical Chemistry Letters **10**, 1942 (2019).
[25] M. Dijkstra, Physical Review Letters **93**, (2004).
[26] A. J. Page and R. P. Sear, Physical Review E **80**, (2009).
[27] E. Sloutskin, et al., Journal of the American Chemical Society **127**, 7796 (2005).
[28] K. S. Gautam and A. Dhinojwala, Physical Review Letters **88**, (2002).
[29] K. S. Gautam et al., Physical Review Letters **90**, (2003).
[30] S. Prasad, Z. Jiang, S. K. Sinha, and A. Dhinojwala, Physical Review Letters **101**, (2008).
[31] M. Ebert, O. Herrmann-Schönherr, J. H. Wendorff, H. Ringsdorf, and P. Tschirner, Liquid Crystals **7**, 63 (1990).
[32] Z. Wu et al., Macromolecules **43**, 4646 (2010).
[33] R. T. Kahl et al., Macromolecules **57**, 5243 (2024).
[34] J. Balko et al., Journal of Materials Research **32**, 1957 (2017).
[35] O. Dolynchuk, et al., Macromolecules **54**, 5429 (2021).
[36] J. Kuebler, T. Loosbrock, J. Strzalka, and L. Fernandez-Ballester, Macromolecules **56**, 3083 (2023).
[37] See Supplemental material at X for details on the sample preparation, measurement setup, data analysis, additional bulk and thin film measurements at room and elevated temperatures, and elaborations on the theory.
[38] R. Xie, R. H. Colby, and E. D. Gomez, Advanced Electronic Materials **4**, (2017).
[39] K. K. Chan et al., Physical Review Letters **54**, 920 (1985).
[40] G. G. Alexander et al., Liquid Crystals **37**, 961 (2010).
[41] K. Binder and P. C. Hohenberg, Physical Review B **6**, 3461 (1972).
[42] E. F. Gramsbergen, L. Longa, and W. H. de Jeu, Physics Reports **135**, 195 (1986).
[43] I. Lelidis and G. Durand, Physical Review E **48**, 3822 (1993).
[44] R. Xie et al., *Macromolecules* 51.24 (2018): 10271-10284.
[45] H. Kasten and G. Strobl. *The Journal of chemical physics* 103.15 (1995): 6768-6774.
[46] T. J. Sluckin and A. Poniewierski, Physical Review Letters **55**, 2907 (1985).
[47] C. Zhao et al., *Nature Communications* 16.1 (2025): 1509.
[48] S. Ren et al. *Molecules* 29.2 (2024): 457.
[49] P. Sheng, Physical Review Letters **37**, 1059 (1976).
[50] S. B. Darling, and M. Sternberg., *The Journal of Physical Chemistry B* 113.18 (2009): 6215-6218.
[51] Z. Cao, et al., *Advanced Materials* 35.41 (2023): 2302178.
[52] Z. Cao et al., Macromolecules **57**, 10379 (2024).
[53] B. I. Halperin, Journal of Statistical Physics **175**, 521 (2019).

*Contact author: oleksandr.dolynchuk@physik.uni-halle.de

# Supplemental Material to “Surface-induced ordering and continuous breaking of translational symmetry in conjugated polymers”

Anton Sinner, Alexander J. Much, Oleksandr Dolynchuk*
Experimental Polymer Physics, Institute of Physics, Martin Luther University Halle-Wittenberg, Halle D-06120, Germany

*Corresponding author: Oleksandr Dolynchuk
Email: oleksandr.dolynchuk@physik.uni-halle.de

**This PDF file includes:**

**with Figures S1-S19, Table S1-S2 and Supporting text**

*Contact author: oleksandr.dolynchuk@physik.uni-halle.de

## S1 Sample preparation, measurement setup and data analysis

**Sample preparation and materials:**

The polymer poly(3-hexylthiophene-2,5-diyl) is a commercially available sample and purchased from BASF and Sigma Aldrich. The BASF batch had a molecular weight of $M_n$=15.6 kg/mol and Đ of 1.6 while the other batch by Sigma Aldrich had a molecular weight of $M_n$=32 kg/mol and DI of 2.15. Both polymers were purified using GPC by Philip Schmode, Uni Bayreuth. Afterwards we had a final molecular weight of $M_n$=15.6 kg/mol and Đ of 1.6 for the BASF sample and $M_n$=47 kg/mol with a Đ of 1.54 for the Sigma Aldrich sample.

The polymer Poly{2,2′-[(2,5-bis(2-hexyldecyl)-3,6-dioxo-2,3,5,6-tetrahydropyrrolo[3,4-c]pyrrole-1,4-diyl)dithiophene]- 5,5′-diyl-alt-thiophen-2,5-diyl} (PDPP[T]$_2$-T) is a commercially available sample with a molecular weight of $M_n$=45.4 kg/mol and Đ of 2.94, purchased from Ossila, Sheffield, England, and was used as received.

All thin films were prepared using the same protocol as described below.

Firstly, substrates were prepared. For this, silicon (Si)-wafers were cut into 8x12mm or 8x15mm substrates. The 8x15mm samples were used such that by removing the edges the area of the 8x12mm and 8x15mm without edges were the same. After cutting the substrates, they were immersed in sulfuric acid to remove any possible organic adhesive and cleaned in petri dishes for 30min. Afterwards, the substrates were rinsed with distilled water and put into a vacuum oven along with a glass bottle for the solution preparation, which was cleaned with ethanol beforehand. The substrates and bottle were heated up to 155°C in vacuum and stayed at the temperature for 30 min and then slowly cooled down. Polymers were dissolved in chloroform in various weight concentrations, heated to 40°C, violently shaken on a vortexer several times, before wrapped into aluminum foil to prevent light interacting with the chain conformation of the semiconducting polymers and transferred to a lab shaker for 60 minutes at 300 motions per minute. Afterwards the substrates from the oven were cleaned with a $CO_2$ snowjet. Substrates were placed onto the spin coated after which 0.75 µl of the solution was pipetted onto the substrate. The samples were spincoated instantly with 2000rpm for 60s. Successively, the short edges, if indicated, where the films are thicker due to material distribution during spincoating were wiped with chloroform to ensure constant film thicknesses during scattering experiments. To check for any major surface impurities after spin coating, all samples were checked under the optical microscope and only samples without major impurities were used.

**Measurement set-up and programs:**

All grazing-incidence wide-angle X-ray scattering (GIWAXS) and WAXS measurements were performed with a laboratory setup Retro-F from SAXSLAB (Copenhagen, Denmark) equipped with a microfocus X-ray sources from AXO (Dresden, Germany) and AXO mulitlayer X-ray optics (ASTIX) as a monochromator for Cu K-α radiation ($\lambda$ = 0.15418nm). A PILATUS R 300K detector from DECTRIS (Daettwil, Switzerland) was used to record 2D scattering patterns. WAXS measurements on bulk samples were performed in transmission geometry and thin films measurements were performed in reflection geometry. For GIWAXS measurements an angle of incidence of $\alpha_i$=0.18° was chosen, in between the critical angle of the substrate ($\alpha_c \sim 0.22°$) and the polymers ($\alpha_c \sim 0.16°$).

All measurements were performed under vacuum and with liquid nitrogen as a coolant on a HFS350 temperature stage, which was connected to a TMS 94 controller and a LNP liquid nitrogen pump, by Linkam Scientific Instruments Ltd., Salfords, UK, with thermal paste to ensure thermal contact, with a sample-detector distance of around 85 mm. For the measurements during melting and crystallization, all P3HT samples were first heated up to a temperature roughly 20°C above the DSC melting peak and equilibrated at that temperature for 30 min and then measured for 30 min to ensure that no peaks were visible. The crystallization was then studied in-situ by GIWAXS starting at ~232°C and then in 1°C steps. Above and below, GIWAXS measurements were performed only at selected temperatures to calibrate the sample position and to save on measurement time. Before every 30 min GIWAXS measurement, a 30 min isothermal step was introduced for P3HT for thermal equilibration. The same measurement protocol was used to study the in-situ melting. Before the appearance and disappearance of the (100) during the crystallization and melting of P3HT, respectively, additional measurements at an incidence angle of $\alpha_i$=2° were introduced. These measurements were performed to check whether the Ewald sphere cut hides the (100) edge-on peak. This was not the case in any of our measurements (exemplary shown in FIG. S2). Thus, in the measured temperature windows, the samples were cooled with a lowest average cooling rate of 1°C/h. A temperature adjusted thermal protocol was used for PDPP[T]$_2$-T. Due to its poorer thermal stability, measurements were shortened such that every isothermal step and measurement were only 10 min long each.

For the isothermal crystallization measurements, the measurements were stopped at the first appearance of the $(100)_{EO}$ peak and the temperature was held isothermally for at least 15 hours.

*Contact author: oleksandr.dolynchuk@physik.uni-halle.de

**Temperature calibration:**

To calibrate the temperature at the surface of the HFS350 temperature stage, a class A PT100 thermocouple was used. The HFS350 temperature was connected to a TMS 94 controller and a LNP liquid nitrogen pump, both by Linkam Scientific Instruments Ltd., Salfords, UK, and operated under vacuum and with liquid nitrogen as a coolant (same measurements set up as above). The PT100 was attached to the temperature stage with a small metal clamp pressing it down. Good thermal contact was further ensured by applying a thin layer of vacuum-suitable thermal grease between the PT100 and the temperature stage. After the temperature stage reached the set temperature according to the internal measurement of the controller, the temperature stage was given 120 seconds to stabilize the temperature. Then the PT100 was read out with a Yocto-PT100 USB temperature sensor by Yoctopuce S.á.r.l., Cartigny, Switzerland. Temperatures were measured in 10°C steps between 20°C and 300°C several times. Before each measurement, the temperature stage and PT100 were cleaned, and fresh thermal grease was used. The internally measured temperature of the Linkam temperature stage were fitted as a function of the temperature measured by the PT100 to determine a calibration function with which all shown temperatures (TMS 94 controller) were shifted to real measured temperatures (PT100).

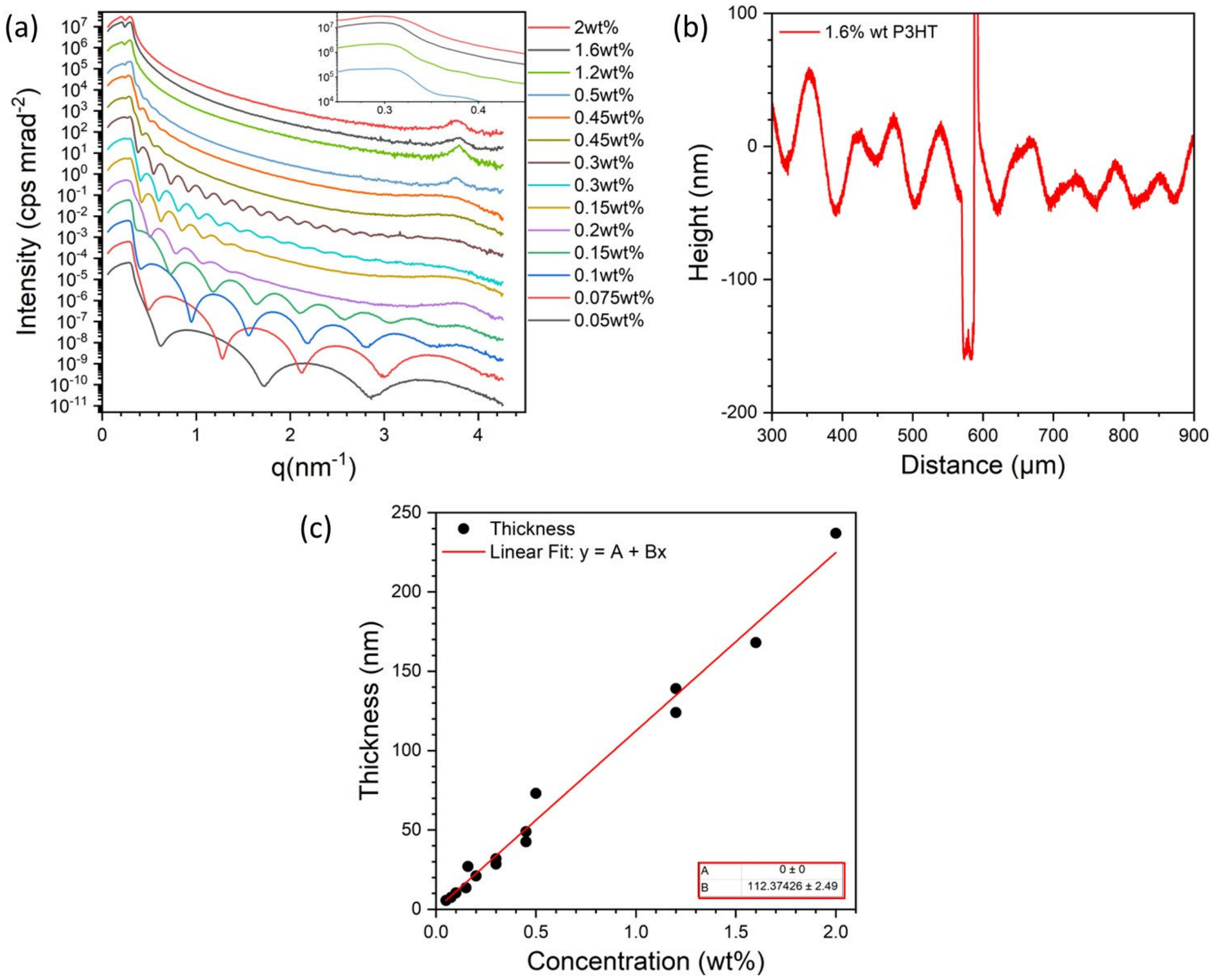

**FIG. S1:** (a) shows representative XRR measurements of P3HT films with $M_n$=15.6 kg/mol on silicon spin coated from solutions with different P3HT concentrations, as indicated in the legend, at 20°C. The curves are vertically shifted to improve the visibility of the Kiessig fringes. The inset enlarges the small q-range to make Kiessig fringes in the 1.2wt% sample better visible. (b) shows a representative profilometer measurement of a scratched P3HT thin film spin coated on silicon with a 1.6 wt% solution. (c) shows the film thickness plotted against the weight concentration of P3HT15 solutions in chloroform used to spin coat the films. The measurements for P3HT47 were performed in the same way and show very similar results.

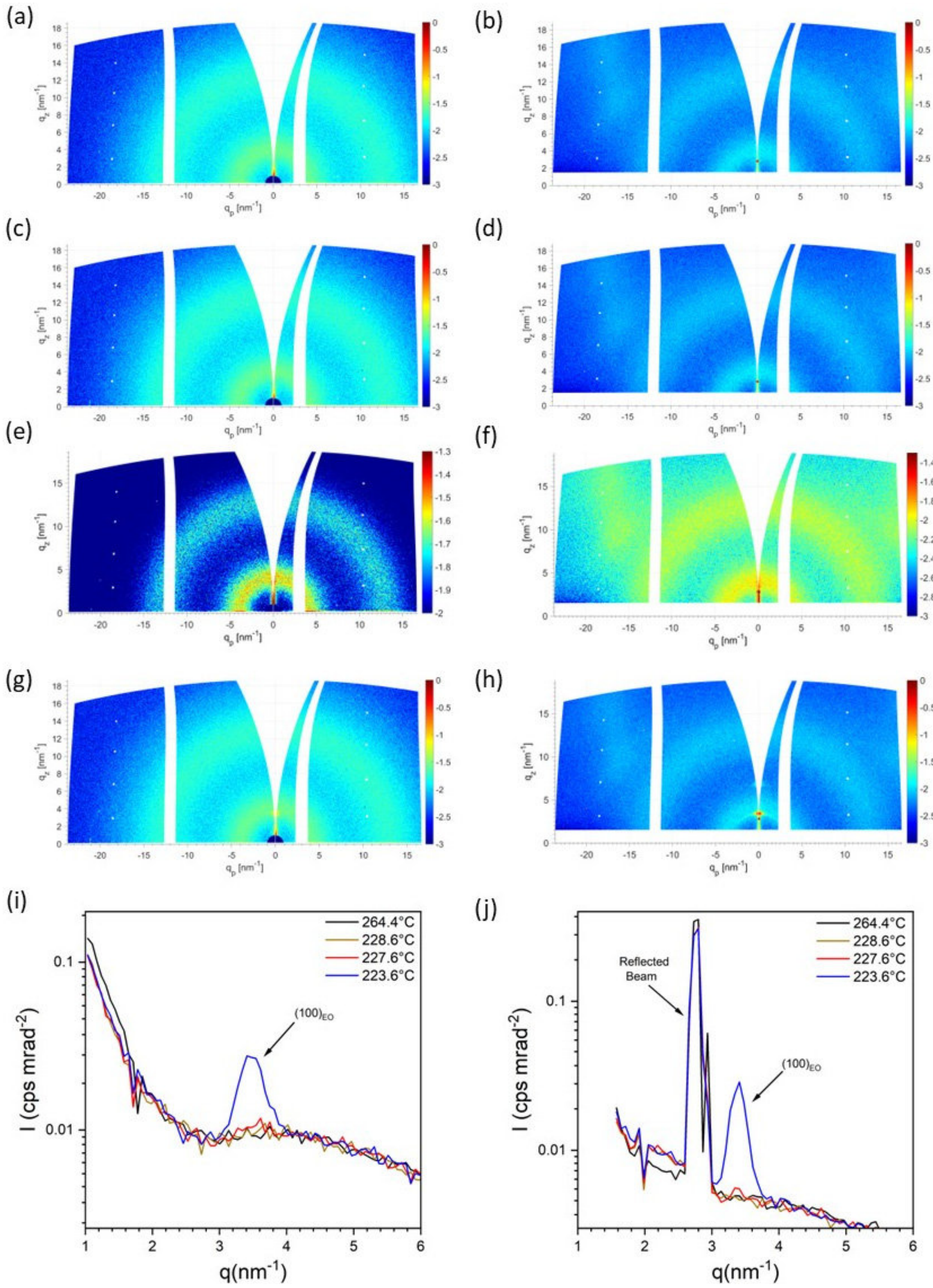

**FIG. S2:** (a)-(h) show exemplary GIWAXS images of a P3HT thin film with a film thickness of 237 nm during cooling at 264.4°C (a, b), 228.6°C (c, d), 227.6°C (e, f) and 223.6°C (g, h) at an incidence angle of $\alpha_i$=0.18° (a, c, e, g) and $\alpha_i$=2° (b, d, f, h). (i, j) show the extracted I(q) curves integrated azimuthally in the range of integration 85-95° for $\alpha_i$=0.18° and $\alpha_i$=2°, respectively. Above 227.6°C an amorphous halo is visible in the GIWAXS images and no peaks are detected, which is supported by the I(q) curves. Starting at 227.6°C an edge-on peak can be detected at both angle of incidence indicating that the Ewald sphere does not hide an earlier appearance of the peak (compare 228.6°C with 227.6°C). By further cooling the sample an increase in the integrated intensity of the edge-on peak is visible.

*Contact author: oleksandr.dolynchuk@physik.uni-halle.de

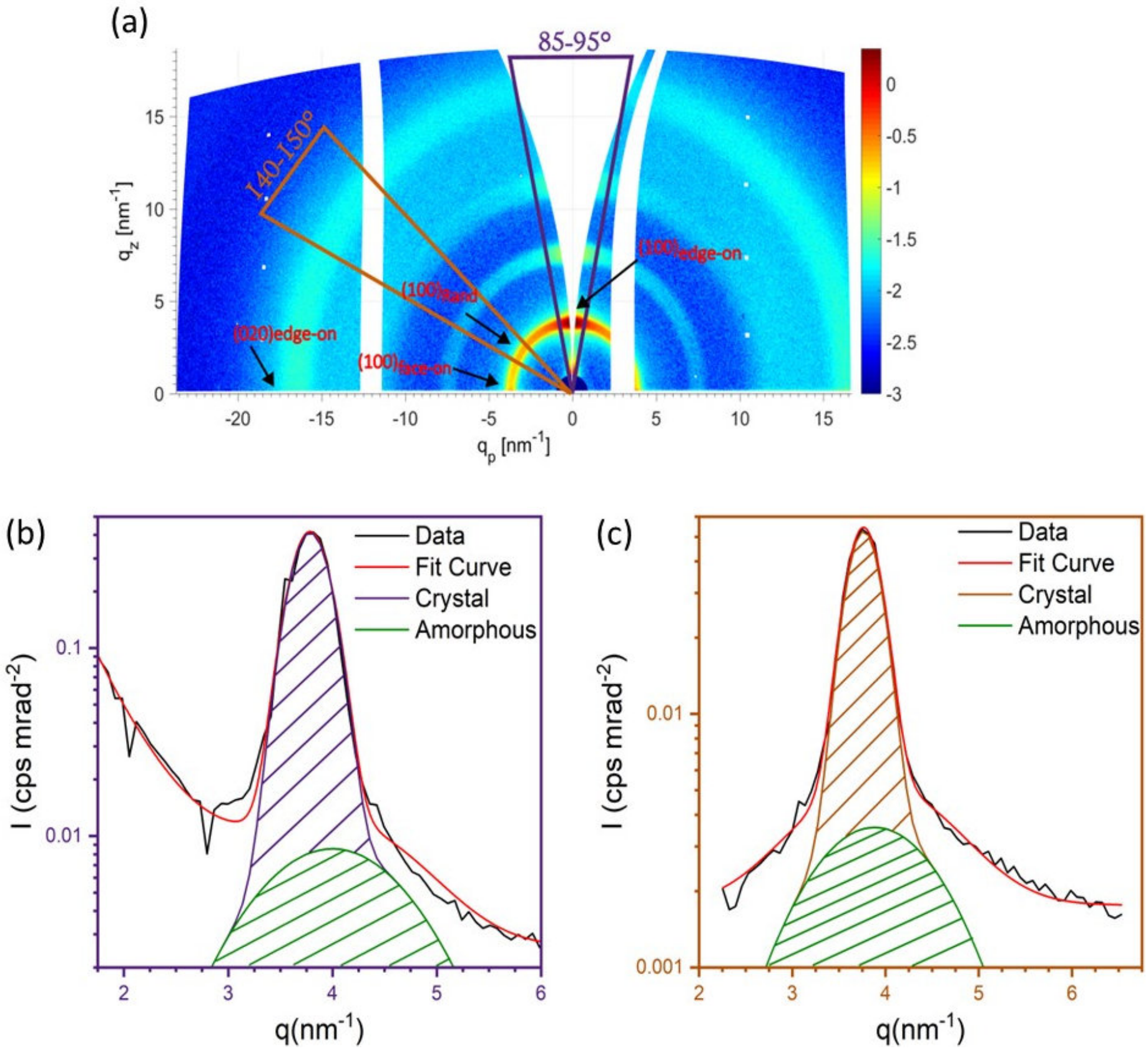

**FIG. S3**: (a) shows a GIWAXS image measured at $\alpha_i$=0.18° of a 237 nm P3HT with $M_n$=15.6 kg/mol thin film after melt crystallization. The (100) and (020) peaks are indicated. The purple and brown triangles schematically show the azimuthal range of integration, namely 85-95° and 140-150°, used to create the one-dimensional curves of I(q). (b) and (c) indicate the 85-95° and 140-150° cut, respectively. The black curve is the azimuthal integrated intensity was fitted with the following function (red curve):

$$I(q) = I_0 + bq^{-n} + \left(\frac{A_{amo}}{w_{amo}\sqrt{\frac{\pi}{2}}}\right) e^{-\left(\frac{q-q_{0,amo}}{w_{amo}}\right)^2} + \left(\frac{A_{cry}}{w_{cry}\sqrt{\frac{\pi}{2}}}\right) e^{-\left(\frac{q-q_{0,cry}}{w_{cry}}\right)^2}, \tag{S1}$$

where $I_0$ is the background, b and n are fitting parameters for the reflected beam for the 85-95° cuts, $A_{amo}$ and $A_{cry}$ is the area of the amorphous and crystalline peak, respectively, $q_{0,amo}$ and $q_{0,cry}$ are the amorphous and crystalline peak position, respectively, and $w_{amo}$ and $w_{cry}$ are the width of the amorphous and crystalline peak, respectively. For the isotropic crystals we set $b$ to zero. Thus, the fitting function takes into account background scattering, the reflected beam, and the contributions from the crystalline peak and amorphous halo. The purple/orange and green colors indicate the crystalline and amorphous contributions, respectively.

*Contact author: oleksandr.dolynchuk@physik.uni-halle.de

| Thickness | Temperature for appearance of $(020)_{EO}$ | Temperature for appearance of $(020)_{Iso}$ |
|---|---|---|
| 237 nm | 220.6°C | 220.6°C |
| 168 nm | 223.6°C | 222.6°C |
| 139 nm | 224.6°C | 220.6°C |
| 73 nm | 222.6°C | Not visible |
| 21 nm | 221.7°C | Not visible |

**Table S1**: The table shows the temperature of appearance for the $(020)_{EO}$ and $(020)_{Iso}$ peak during the stepwise cooling for P3HT with $M_n$=15.6 kg/mol. Since crystallization is a non-equilibrium process, the varying temperatures for the appearance of the $(020)_{EO}$ are to be expected. However, it becomes apparent that only for the thickest film the appearance of the $(020)_{EO}$ and $(020)_{Iso}$ peak happen at the same temperature. With decreasing film thickness the temperature of appearance for the $(020)_{Iso}$ peak decreases until it is not seen for the thinnest films. Since the intensity of the $(020)_{Iso}$ peak is distributed on a circle on the Ewald-sphere and with decreasing film thickness the amount of isotropic liquid crystals decreases, it is likely that this effect is due to the limit of the measurement sensitivity.

*Contact author: oleksandr.dolynchuk@physik.uni-halle.de

## S2 WAXS and DSC measurements of bulk P3HT

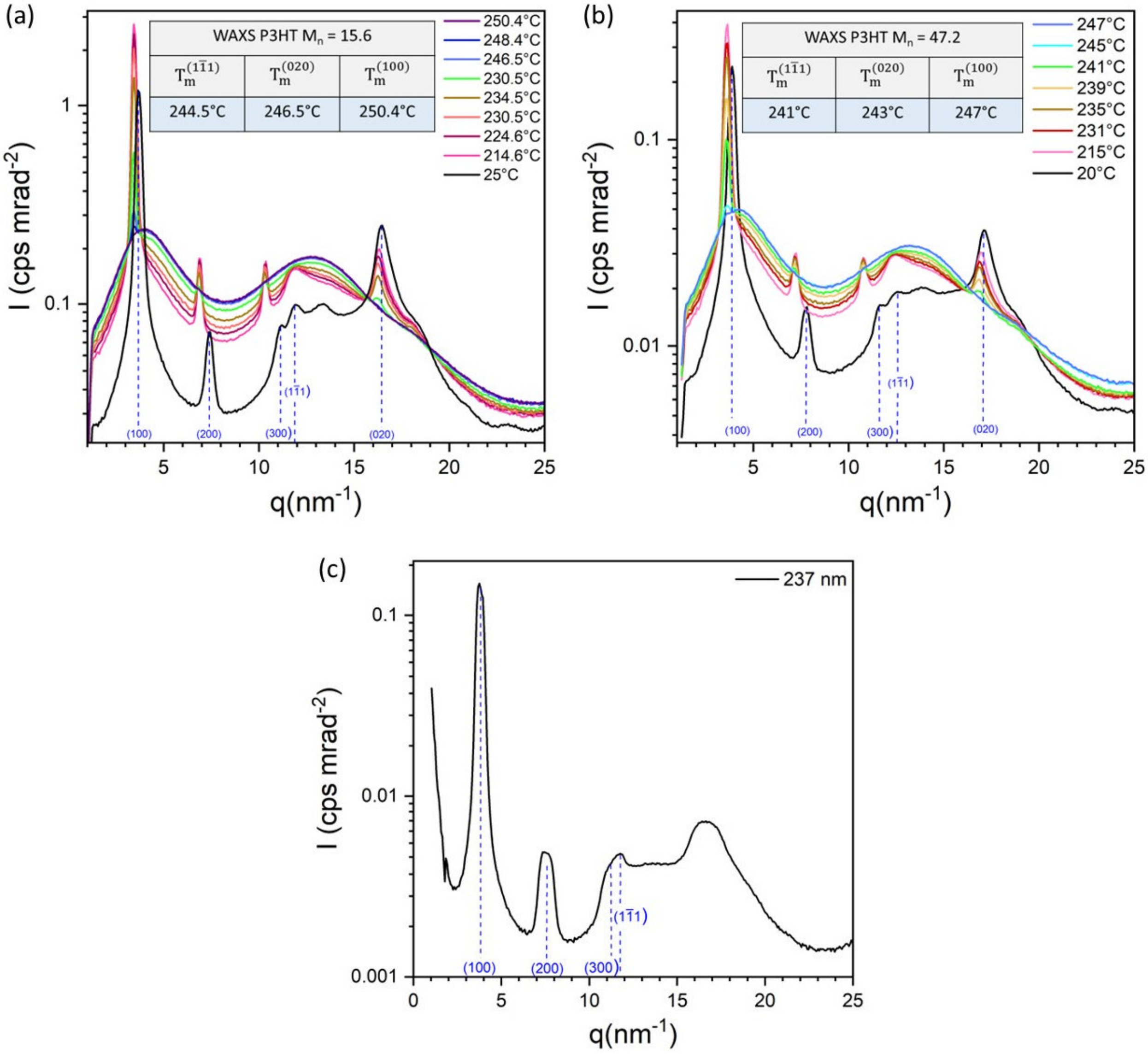

**FIG. S4:** The intensity plotted against the scattering vector for bulk P3HT with $M_n$=15.6 (a) and $M_n$=47.2 (b) kg/mol measured in transmission geometry at different temperatures during heating after the sample crystallized from the melt at 1 K/min. Tables showing the respective melting temperatures at which the various peaks were no longer visible are included. (c) shows a polar integration in the azimuthal range 0 to 180° for a P3HT thin film with $M_n$=15.6 kg/mol and a thickness of 237 nm after melt crystallization. Indicated are all visible reflections including the $(1\bar{1}1)$ reflection.

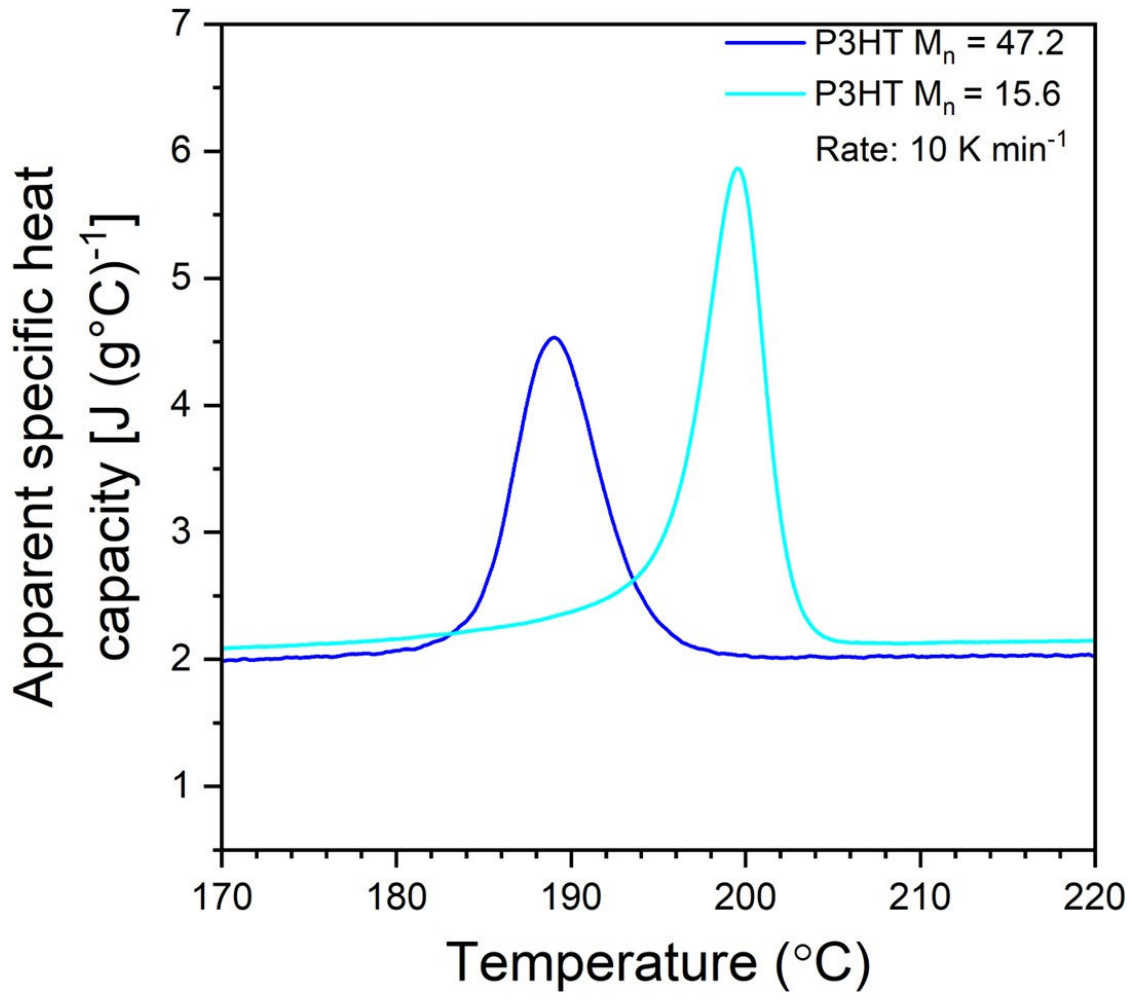

**FIG. S5:** Apparent specific heat capacity measured of P3HT with $M_n$=15.6 (cyan) and $M_n$ = 47.2 (blue) against temperature during cooling with a rate of 10 K/min.

# S3 GIWAXS measurements on thin films at room temperature

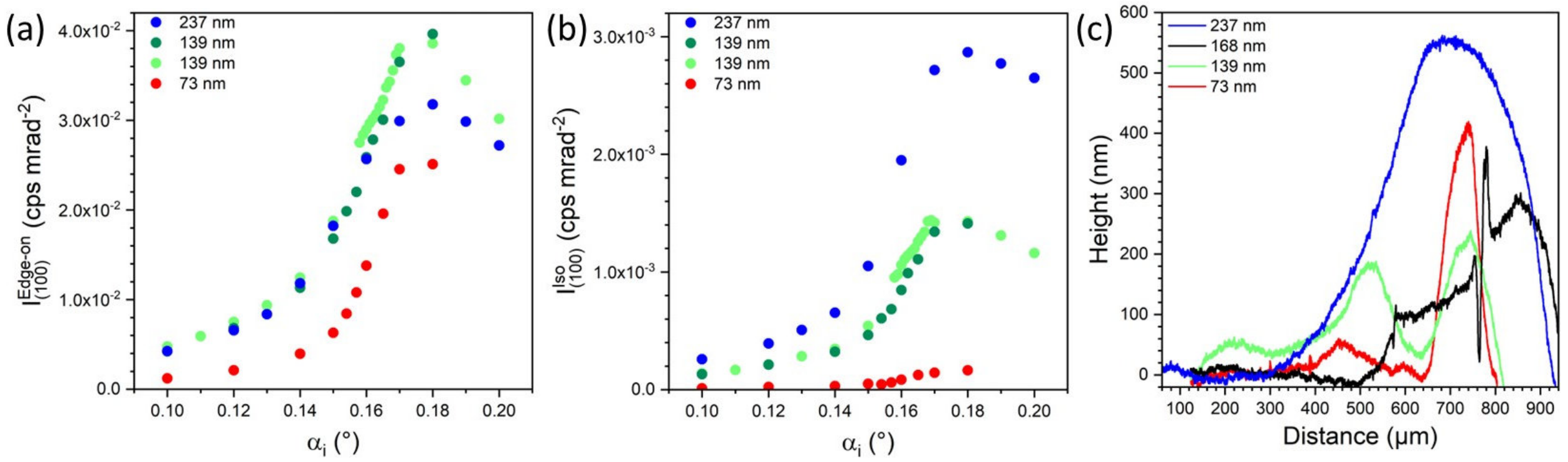

**FIG. S6:** (a) Intensity $I_{(100)}^{Edge-on}$ for the edge-on oriented crystals and (b)intensity $I_{(100)}^{Iso}$ for the isotropic crystals versus the angle of incidence, $\alpha_i$, at room temperature for P3HT films with $M_n$=15.6 kg/mol with various film thicknesses, indicated in the legend, after crystallization from the melt. By decreasing the angle of incidence below the critical angle of P3HT ($\alpha_i$=0.163°), varying penetration depths can be probed. With decreasing angle of incidence, the penetration depth decreases, which allows us to probe only the top surface. At $\alpha_i$=0.18°, where the whole film thickness is penetrated, a clear non monotonic behavior is visible for $I_{(100)}^{Edge-on}$ in comparison to a monotonic increase in $I_{(100)}^{Iso}$. Since with increasing film thickness the amount of isotropic crystals increases linearly, a monotonic behavior for the edge-on oriented crystals should also be expected. Thus we checked the edges of the sampel with profilometer measurements (c), where a tip is dragged across the edge of the sample and the height profile is measured. A clear increase in the height of the edges is visible. This leads us to conclude that the edges have a sizeable impact on the surface sensitive GIWAXS measurements. In Fig. S7, the same measurements for the same samples with removed edges are shown. The behavior of $I_{(100)}^{Edge-on}$ is more sensible.

*Contact author: oleksandr.dolynchuk@physik.uni-halle.de

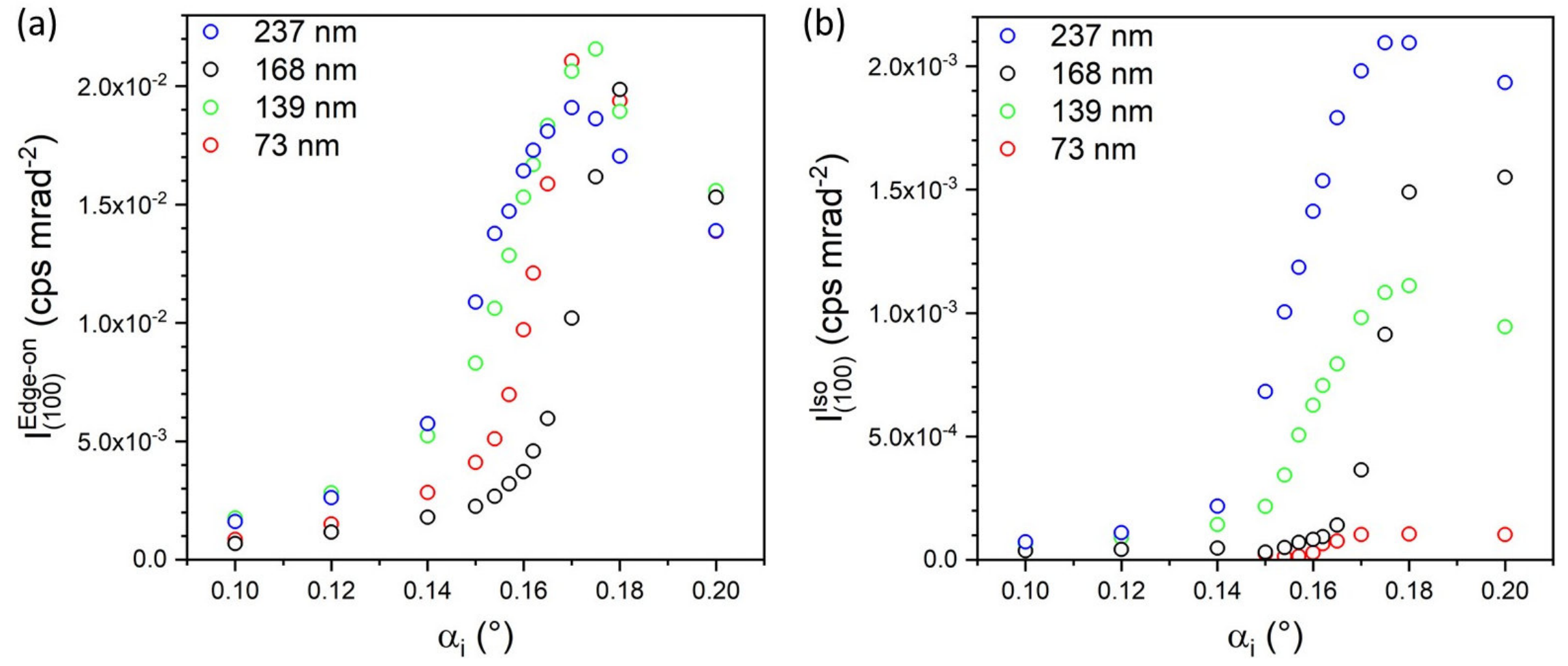

**FIG. S7:** (a) Intensity $I_{(100)}^{Edge-on}$ for the edge-on oriented crystals and (b)intensity $I_{(100)}^{Iso}$ for the isotropic crystals versus the angle of incidence, $\alpha_i$, at room temperature for P3HT films with $M_n$=15.6 kg/mol with various film thicknesses, indicated in the legend, after crystallization from the melt. By decreasing the angle of incidence below the critical angle of P3HT ($\alpha_c$=0.163°), varying penetration depths can be probed. With decreasing angle of incidence, the penetration depth decreases, which allows us to probe only the top surface. For all samples the edges were removed which leads to an overlap for $I_{(100)}^{Edge-on}$for all film thicknesses and a monotonic increase for $I_{(100)}^{Iso}$ at $\alpha_i$ = 0.18°, indicating that while the amount of edge-on oriented crystals stays constant the amount of isotropic crystals increases monotonically with film thickness. Furthermore $I_{(100)}^{Iso}$ drops of quicker than $I_{(100)}^{Edge-on}$ indicating that the edge-on oriented crystals are above the isotropic crystals. Specifically, for the 73 nm, there is no intensity contribution for lower $\alpha_i$ and a marginal contribution for the other film thicknesses. This marginal contribution is likely due to the surface roughness of the film, which locally changes the incidence angle and thus the final penetration depth.

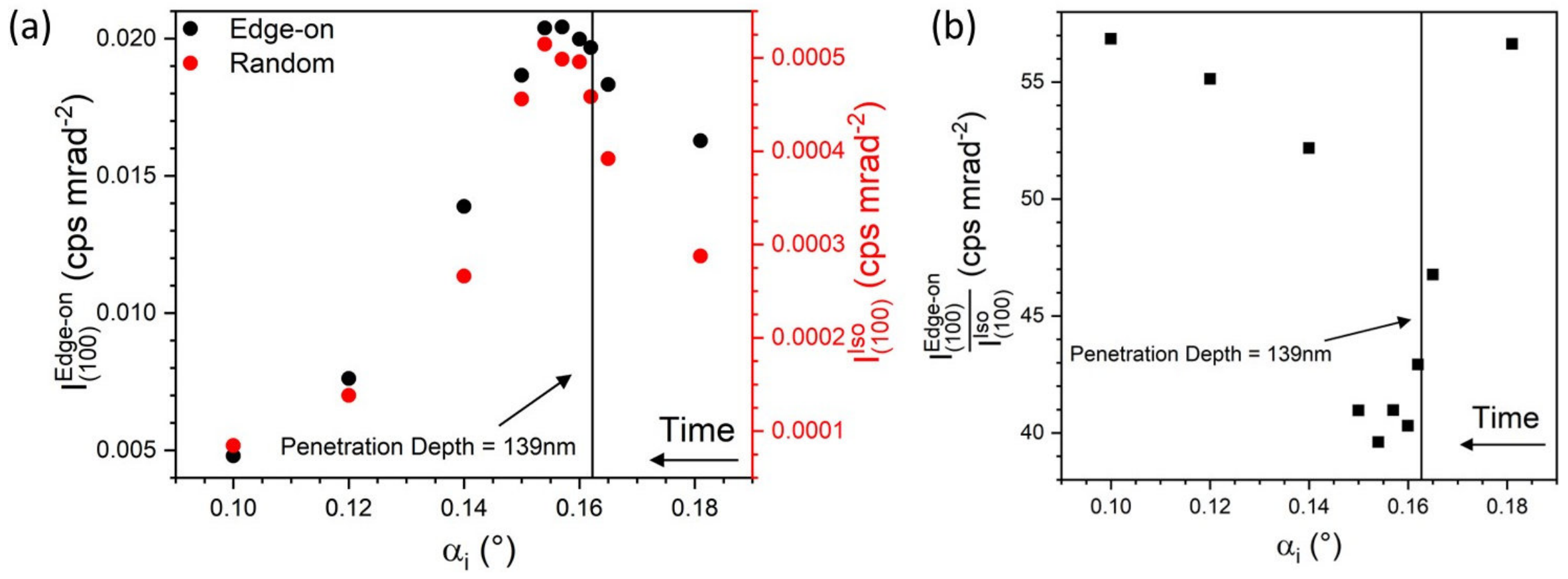

**FIG. S8:** In (a) the integrated intensity for the edge-on oriented (black) and isotropic (red) LCs is measured at various angles of incidences, $\alpha_i$, at ~227°C, for a P3HT film with a thickness of 139 nm with $M_n$=15.6 kg/mol. By decreasing the angle of incidence below the critical angle of P3HT ($\alpha_c$=0.163°), varying penetration depths can be probed. With decreasing angle of incidence, the penetration depth decreases, which allows us to probe only the top surface. Every measurement took 60 minutes. In figure b), the ratio between the edge-on and isotropic liquid crystals is plotted against the incidence angle. We can clearly see at first a decrease in the ratio which indicates that the isotropic LCs grow quicker than the edge-on oriented liquid crystals. Below an incidence angle of $\alpha_i$=0.15, we can see an increase in the ratio. Below $\alpha_i$=0.15, the x-rays start to penetrate less deeply and only the top surface is scattered. Thus, an increase in the ratio suggests that we have more edge-on oriented LCs at the top surface.

*Contact author: oleksandr.dolynchuk@physik.uni-halle.de

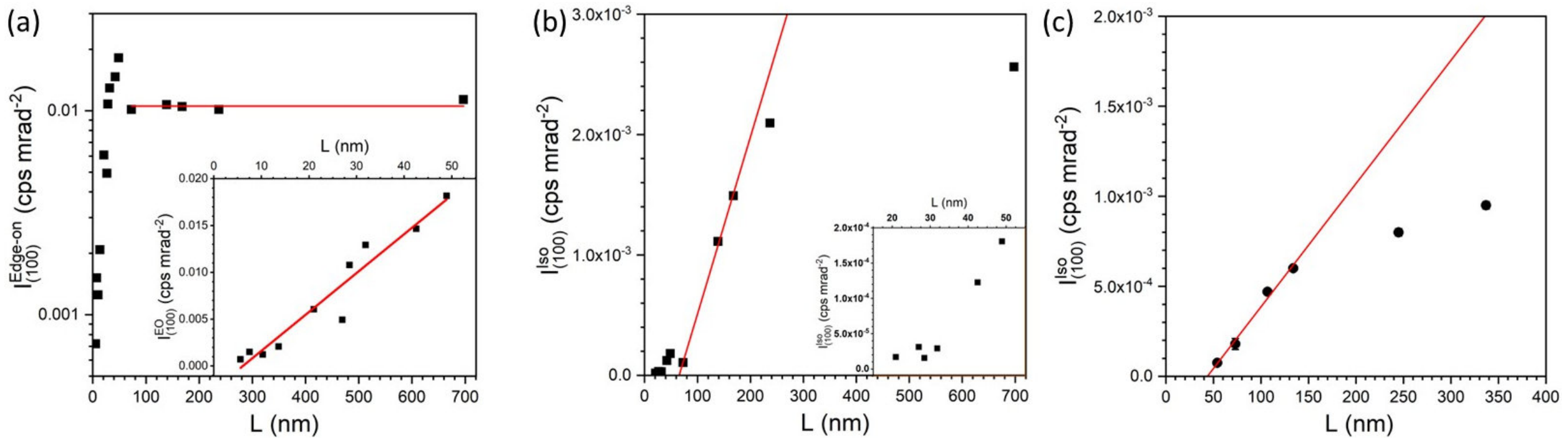

**FIG. S9**: (a) Intensity $I_{(100)}^{Edge-on}$ versus film thickness for P3HT films with $M_n$=15.6 kg/mol. (b) and (c) intensity $I_{(100)}^{Iso}$ versus film thickness for P3HT films with $M_n$=15.6 kg/mol and $M_n$=47.2 kg/mol, respectively. The samples in (a), (b) and (c) were measured after cooling from the melt with 2 °C/min without edges. In (a) two regimes can be identified. Below 73 nm we see a linear increase in $I_{(100)}^{Edge-on}$ with film thickness while for 73nm and above we see a plateau in $I_{(100)}^{Edge-on}$. This suggests that the edge-on layer increases linearly until a specific thickness where it plateaus. Looking at (b) until 35 nm we see a constancy in isotropic crystals after which we see a slight increase in intensity up to 73 nm. Then starting at 73 nm, we see a linear behavior in $I_{(100)}^{Iso}$ with film thickness which starts to deviate starting at roughly 200 nm. This deviation is probably due to absorption effects. A qualitative similar behavior is observed in (c). Both are fitted with a line (red). These observations lead us to several conclusions. For films above 73 nm, the constancy in $I_{(100)}^{Edge-on}$ (a) suggests that the thickness of the edge-on layer stays constant. If the thickness of the edge-on layer stays constant with increasing thickness and $I_{(100)}^{Iso}$ shows a linear behavior up to 200 nm, we can extrapolate this linear behavior to zero $I_{(100)}^{Iso}$ to get the thickness of the edge-on layer. By extrapolating to zero we get a thickness for the edge-on layer of 60 nm for P3HT with $M_n$=15.6 kg/mol and 40 nm for P3HT with $M_n$=47.2 kg/mol. However, looking closely at (a) and (b), we can see that at 50 nm $I_{(100)}^{Edge-on}$ has a larger value than all the higher film thicknesses, suggesting a thickness below 60 nm. Furthermore, we still observe isotropic crystals below 60 nm (b). Since the films are very thin and as seen in Fig. S13, the silicon interface plays a role in the melting of P3HT with $M_n$=15.6 kg/mol, we presume that the influence of the silicon interface cannot be ignored anymore either due to confinement effects or surface melting. Furthermore, the kinematic approximation for scattering is probably not accurate enough and the Distorted Wave Born approximation must be taken into account. This leads to distortions in the measured intensity.

*Contact author: oleksandr.dolynchuk@physik.uni-halle.de

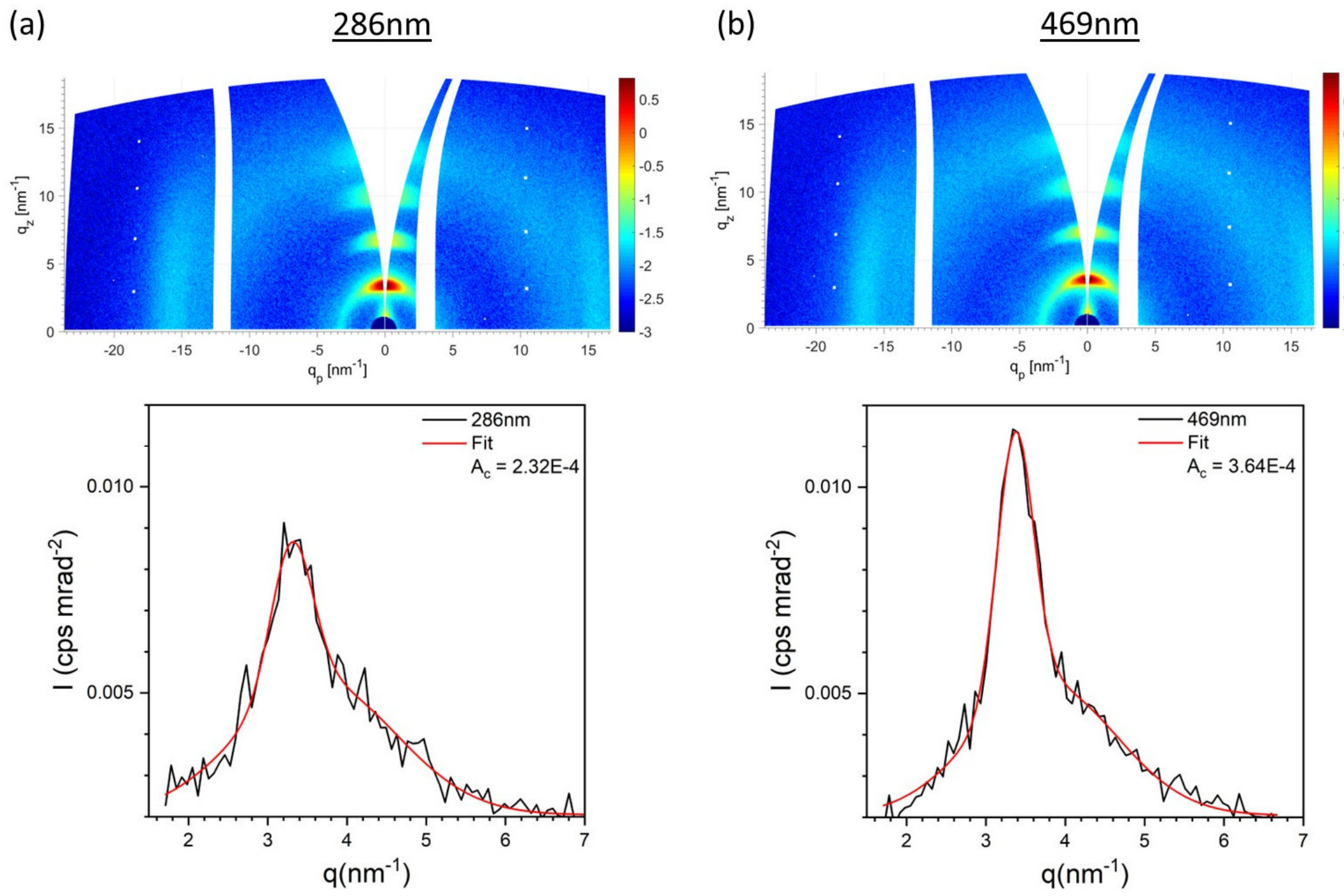

**FIG. S10:** a) and b) show the GIWAXS images and azimuthally integrated intensity in the range between 140-150° versus q for PDPP[T]$_2$-T for two different film thicknesses. Although isotropic crystals are not visible in the GIWAXS image due to their low intensity, they can be identified in the 140-150° line cuts. The integrated intensity can be fitted with Eq. S1 by setting $b = 0$, and the crystalline area can be extracted.

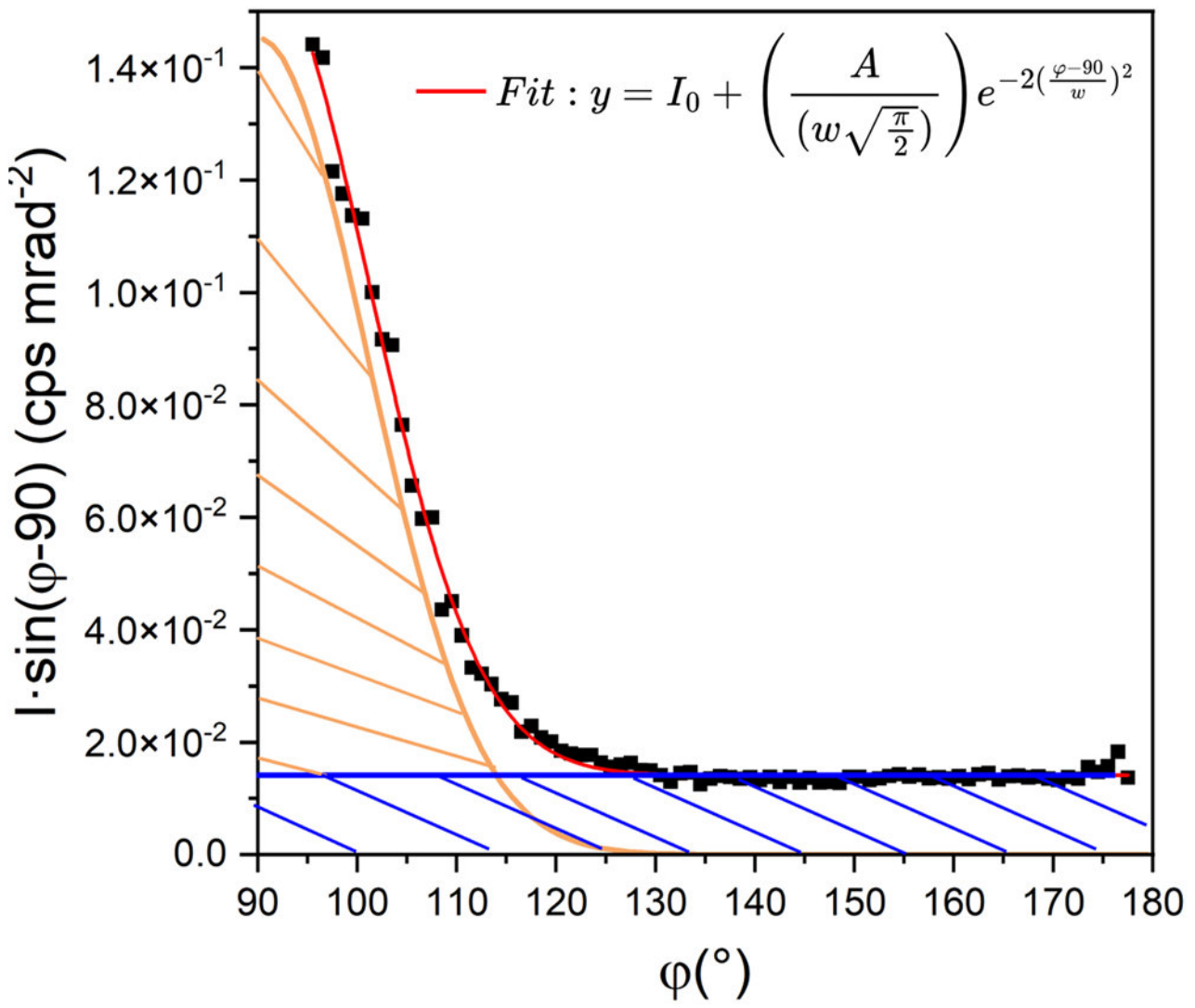

**FIG. S11:** The sin(φ) corrected azimuthal intensity (black squares) of the (100) peak versus azimuthal angle, φ, is depicted for a PDPP[T]$_2$-T film with a thickness of 286 nm. We fit (red curve) the dependence by using a Gaussian function for the edge-on oriented liquid crystals (orange curve) and a constant background for the isotropic liquid crystals (blue curve). By taking the ratio between the area under the Gaussian peak and the total fitted intensity, we get the fraction of edge-on oriented liquid crystals given as: $\frac{A}{(180*I_0+A)}$. By multiplying this ratio with the film thickness, we obtain an estimate for the thickness for the edge-on oriented layer of about 180 nm.

*Contact author: oleksandr.dolynchuk@physik.uni-halle.de

## S4 GIWAXS measurements on thin P3HT films at high temperatures

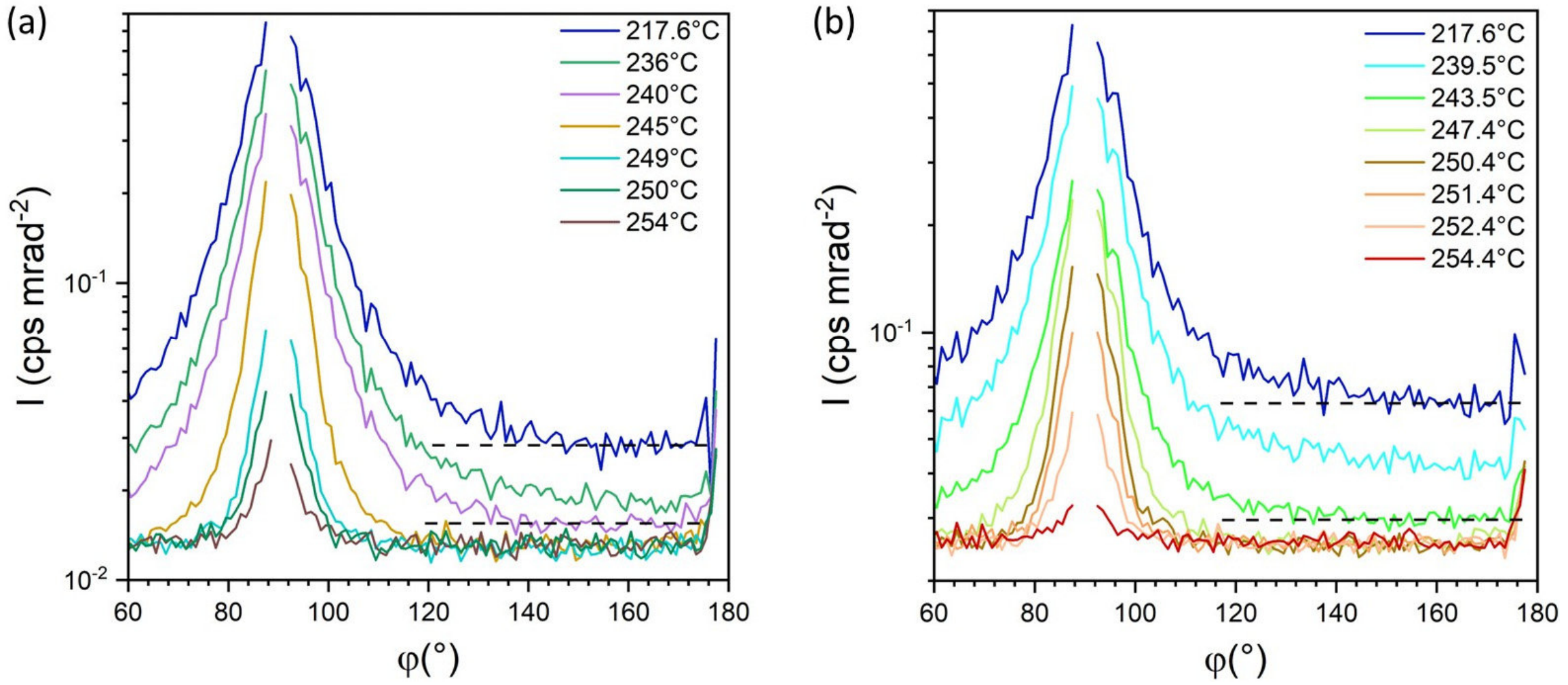

**FIG. S12:** Pole figures for the (100) peak for the 139nm (a) and 237nm (b) thick P3HT films with $M_n$=15.6 kg/mol during heating, respectively. During heating the width of the azimuthal distribution of edge-on oriented LCs decreases, suggesting that nearly tilted edge-on oriented crystals melt first. Dotted lines included as guide to eye.

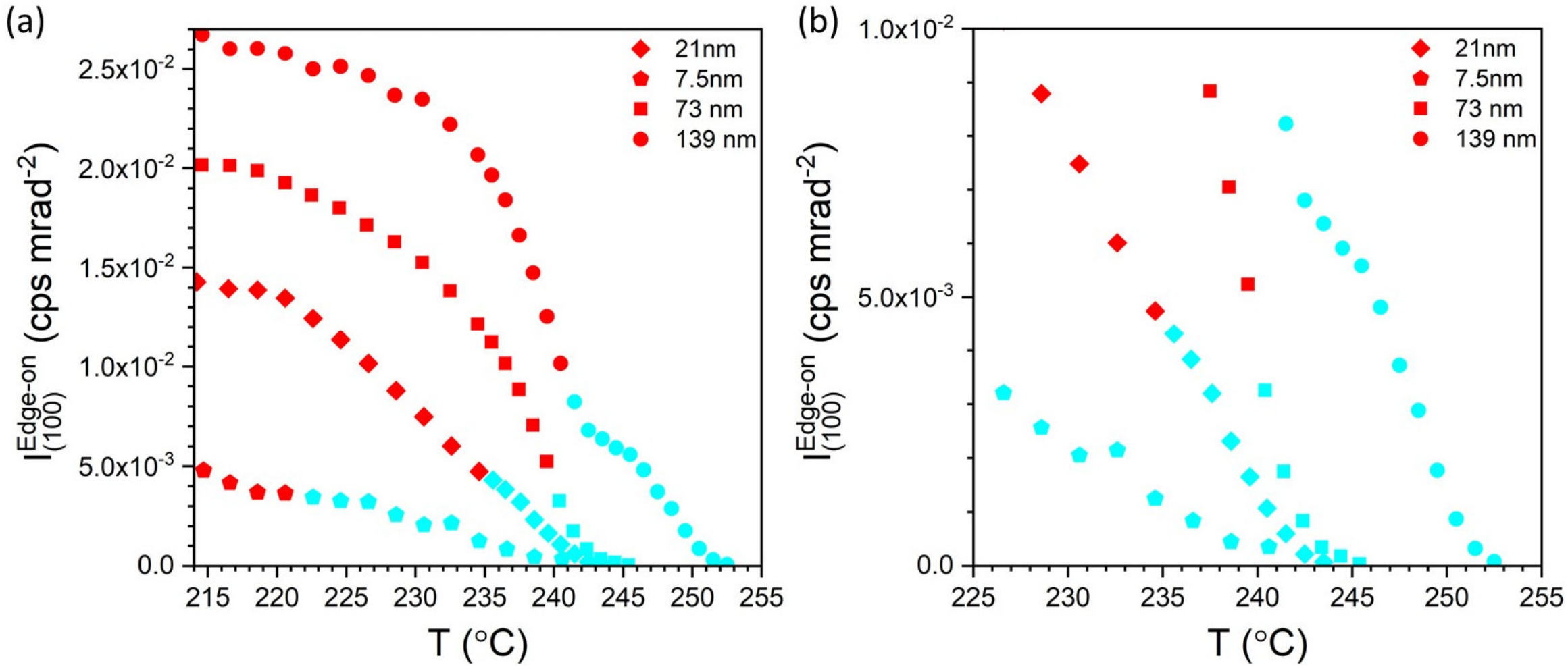

**FIG. S13:** (a) Intensity $I_{(100)}^{Edge-on}$ for edge-on oriented LCs as a function of temperature of P3HT films with a thickness below 139 nm after stepwise cooling; (b) enlarges the region of high temperature in (a). Cyan and red indicate a $\Sigma_d$ - and $\Sigma_o$ LC state, respectively. The disappearance of the quasi-plateau is observed for film thicknesses below 139 nm. Furthermore, a monotonic decrease in melting temperature and $\Sigma_o$ to $\Sigma_d$ transition temperature with decreasing film thickness is observed. This indicates that for thinner films the silicon interface has a strong effect on the melting behavior, suggesting either confinement effects or surface melting at the silicon interface.

*Contact author: oleksandr.dolynchuk@physik.uni-halle.de

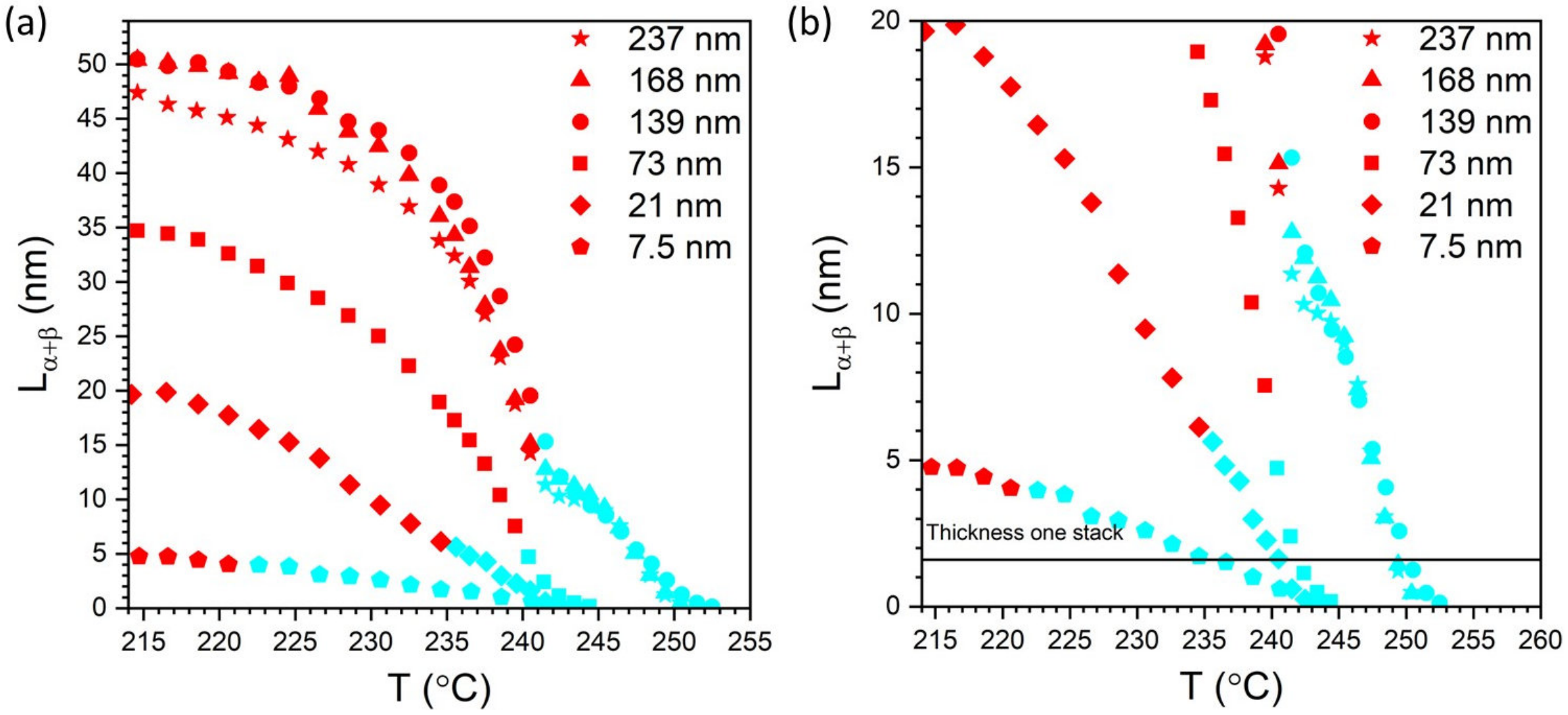

**FIG. S14:** (a) Shows the estimate of the thickness of the edge-on oriented layer for P3HT with $M_n$=15.6 kg/mol films on silicon for various thicknesses, as indicated in the legend. (b) enlarges the high temperature region of (a) The thinnest film with $L$ = 7.5nm has only edge-on crystals and the measured intensity was taken to estimate the thickness of one stack. This intensity was used to turn the measured intensity $I_{(100)}^{Edge-on}$ into thickness. Note that in the entire calculation the form factor is assumed to be temperature independent, thus only the number of (100) layers are temperature dependent. However, it leads to sensible film thicknesses as can be seen in (a) and (b). The alpha layer can then be estimated to have a thickness of roughly 12 nm which would be six layers. However, as can be seen in (b) this makes two facts apparent. First, we do not have a discrete decrease in thickness, which would be expected if the whole lateral region, meaning a whole layer, melts instantaneously. Secondly, we apparently measure a thickness below one stack which is not possible in scattering, suggesting again non uniform melting in the lateral planes. Lastly, the 7.5 nm film has a specific (100) azimuthal distribution, which we expect to scale linearly with film thickness. This also leads to deviations in the thickness.

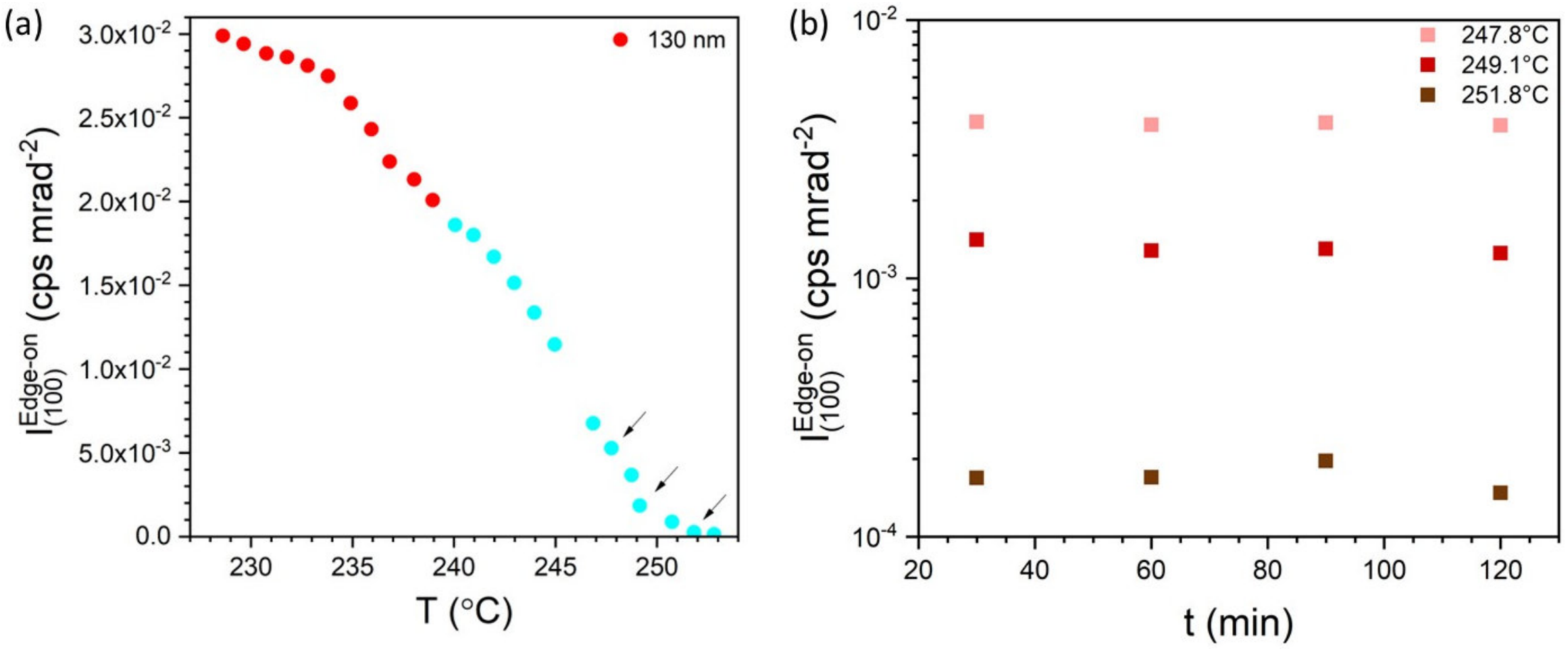

**FIG. S15:** (a) Intensity $I_{(100)}^{Edge-on}$ versus temperature during heating of P3HT film on silicon ($L$≈130 nm) with $M_n$=15.6 kg/mol after isothermal crystallization. Arrows indicate at which points the isothermal measurements were introduced. (b) Time evolution of the intensity $I_{(100)}^{Edge-on}$ in the $\Sigma_d$ -LC state, where no isotropic LCs are observed. Cyan and red indicate a $\Sigma_d$ - and $\Sigma_o$ LC state, respectively. At three different temperatures four measurements, each 30 minutes, were done. At all three temperatures $I_{(100)}^{Edge-on}$ show barely any variation indicating that melting is an equilibrium process.

## S5 AFM and POM measurements on P3HT films

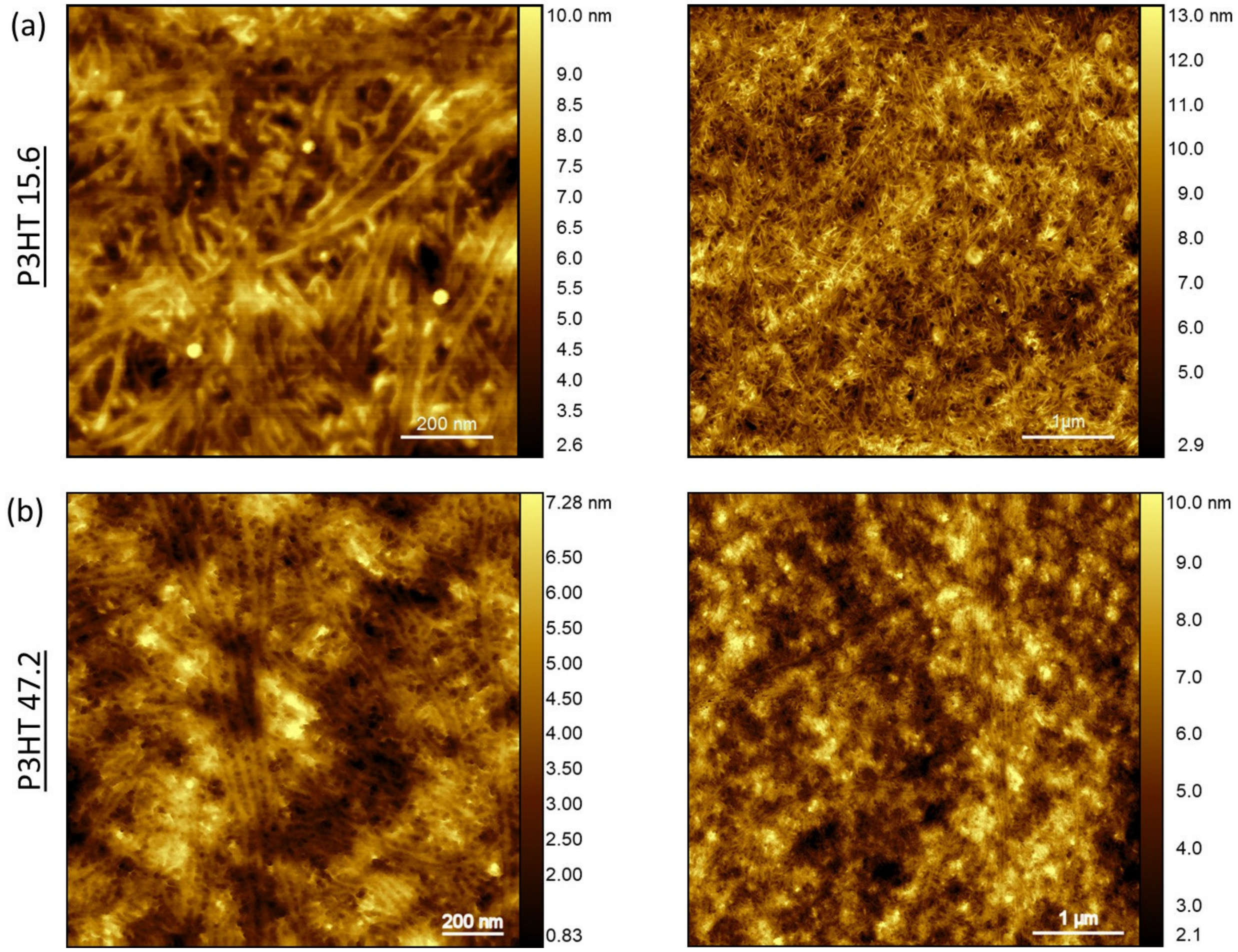

**FIG. S16:** a) and b) show representative AFM height images of P3HT films with a film thickness of roughly 140 nm with a molecular weight of $M_n$=15.6 and $M_n$=47.2 kg/mol after slow crystallization from the melt. A typical lamella structure for both molecular weights can be seen, although it is better visible for the lower molecular weight, which is to be expected due to its higher crystallinity. There appear some points where multiple lamellae seem to cross or meet however no spherulites were found.

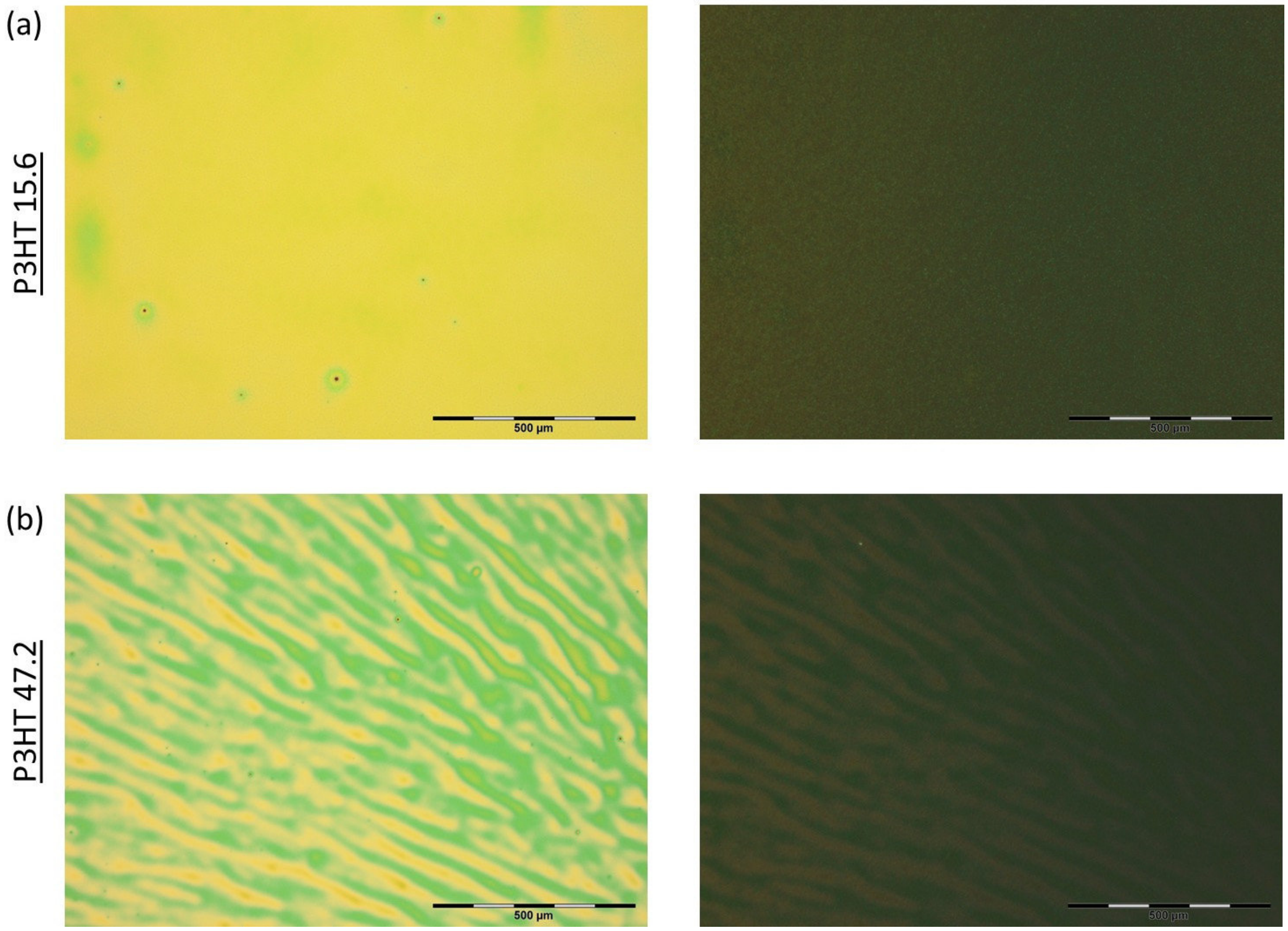

**FIG. S17:** a) and b) show representative optical microscopy images under open (left column) and 90°-crossed (right column) polarizers of P3HT films with a molecular weight of $M_n$=15.6 (a) and $M_n$=47.2 kg/mol (b) with a thickness of about 140 nm after slow cooling from the melt. A small birefringent signal is visible for the lower-$M_n$ sample while for the higher- $M_n$ sample no signal is seen. Furthermore, while for the lower-$M_n$ sample the Marangoni patterns vanish after crystallization from the melt, for the higher- $M_n$ sample they are still present.

*Contact author: oleksandr.dolynchuk@physik.uni-halle.de

## S6 Landau type theory of surface freezing

**Weak first-order phase transition**

In the 1930s, Landau, and later by collaboration with Ginzburg, developed a mean field theory to describe phase transitions which connects the symmetry breaking of the system to the order parameter. They made the simple assumption that the free energy density can be approximated as analytic function close to the phase transition and can be expressed as an expansion of the free energy density in terms of powers in the order parameter, since the later changes from zero to some value at the transition. Depending on the order parameter of the system, various terms are forbidden due to symmetry considerations. The free energy density for a weak first-order phase transition with an order parameter, $\eta$, is given by:

$$f(\eta, T) = f_0 + \frac{1}{2}a\eta^2 - \frac{1}{3}b\eta^3 + \frac{1}{4}c\eta^4, \tag{S2}$$

where, $a = a_o(T - T_0)$ with $T_0$ being the temperature, where one global minimum exist which is non zero, and $a_0$, $b$ and $c$ being phenomenological parameters. By minimizing the free energy, the order parameter profile can be derived as:

$$\eta = \frac{\tilde{b}}{2} + \sqrt{\frac{\tilde{b}^2}{4} - \tilde{a}}, \tag{S3}$$

where $\tilde{a} = \frac{a}{c} = \frac{a_0}{c}(T - T_0) = \tilde{a}_0(T - T_0)$ and $\tilde{b} = \frac{b}{c}$. The melting temperature $T_m$ can be derived from these expressions by inserting the order parameter from Eq. S3 into the free energy from Eq. S2 and calculating the temperature at which the minimum of the free energy difference, $f(\eta, T) - f_0$, reaches zero. This leads to $T_m = T_0 + \frac{2\tilde{b}^2}{9\tilde{a}_0}$. With this the order parameter jump at $T_m$ can be easily calculate with Eq. S2 which is $\eta(T_m) = \frac{2\tilde{b}}{3}$. Lastly, the maximum supercooling is obtained by subtracting $T_0$ from the temperature $T^*$, at which a second minima vanishes. This leads to a maximum supercooling of $\Delta T_{max} = T^* - T_0 = \frac{\tilde{b}^2}{4\tilde{a}_0}$.

**Surface freezing as introduced by Lipowsky**

In 1982, Lipowsky introduced a framework to treat surface induced melting and freezing as an analogy to the semi-infinite Ising model [14,41].

We consider a three-dimensional system in which the z-direction is perpendicular to the surface and the x- and y-direction span the surface. The surface plane is at $z = 0$. To the standard Landau free energy density, we now add a coupling term $\frac{1}{2}J(\nabla\Psi)^2$ describing the fluctuation in the order parameter and we assume that the surface has an energetic contribution $f_s(\Psi)$ distinct from the bulk free energy. Lastly, the surface breaks the translational symmetry in the z-direction, thereby making the order parameter, $\Psi$, a z-dependent function, $\Psi(z)$. Thus, the free energy density is given by:

$$F(\Psi, T) = \int_0^\infty dz \left[\frac{1}{2}J\big(\nabla\Psi(z)\big)^2 + f\big(\Psi(z)\big) + \delta(z) f_s\big(\Psi(z)\big)\right], \tag{S4}$$

where $F(\Psi, T)$ is the free energy per unit area of the whole system, $f\big(\Psi(z)\big)$ and $f_s\big(\Psi(z)\big)$ are the free energy per unit volume of the bulk and the surface, respectively. By the variational principle $\delta F = 0$, analogously to Euler-Lagrange theory, the surface- and bulk order parameter can be derived which are given as:

$$\left.\frac{\partial f_s}{\partial \Psi}\right|_{z=0} = J \left.\frac{\partial \Psi}{\partial z}\right|_{z=0} \text{ and } \left.\left(\frac{\partial \Psi}{\partial z}\right)\right|_{z=0} = \begin{cases} +\left(\frac{f(\Psi) - f(\Psi_b)}{J}\right)^{\frac{1}{2}}, \Psi_s < \Psi_b, \\ -\left(\frac{f(\Psi) - f(\Psi_b)}{J}\right)^{\frac{1}{2}}, \Psi_s > \Psi_b, \end{cases} \tag{S5}$$

where the first expression is a boundary condition coming from minimizing $\delta F$ and the second expression can be derived from the Euler equation. $\Psi_s$ and $\Psi_b$ are the order parameter values at the surface $(z = 0)$ and in the bulk, respectively. Since we consider surface ordering, we have as a condition $\Psi_s > \Psi_b$ and $\Psi_b = 0$, because the bulk, in our case, is in an unordered state.

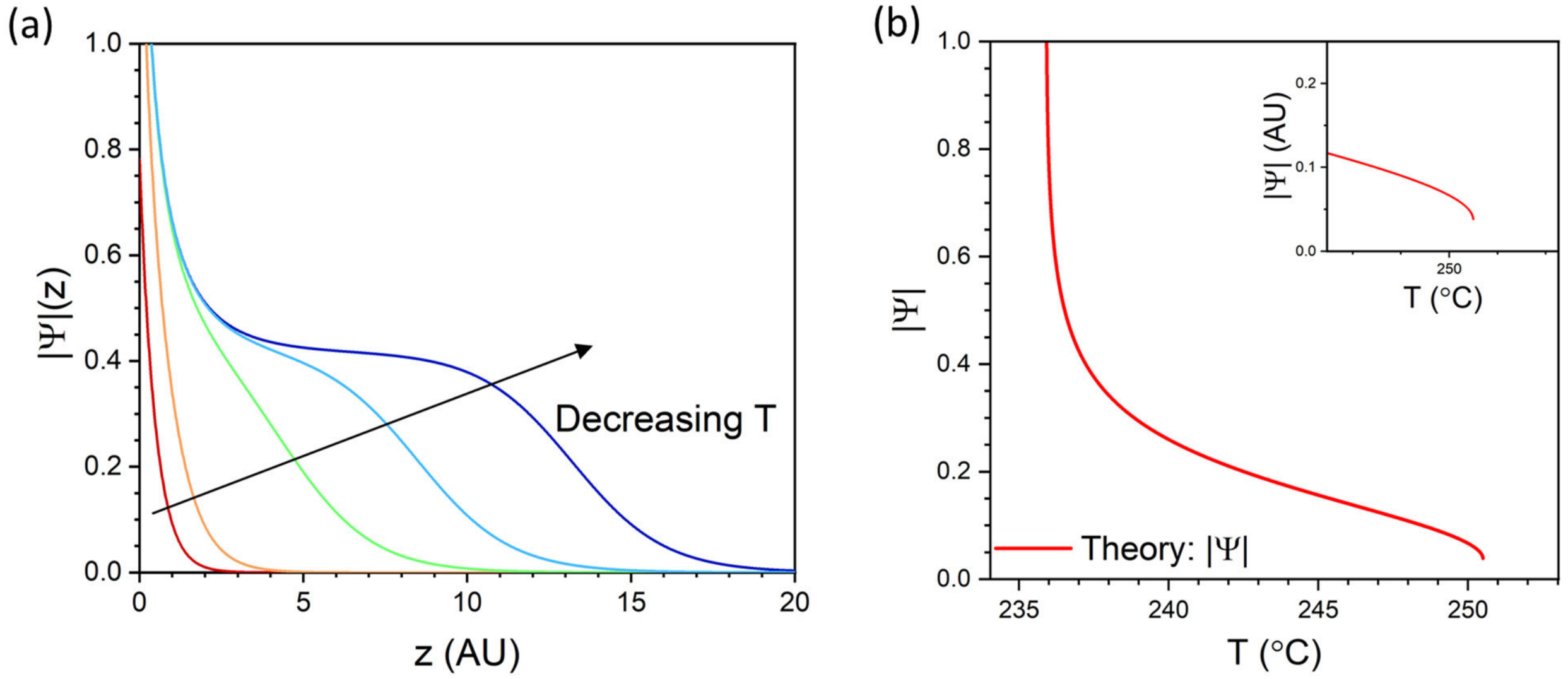

**Fig. S18**: (a) The order parameter $|\Psi|(z)$ (Eq. S7) plotted against the depth, z, for varying temperatures from highest (red) to lowest (blue). The ordering starts at the vacuum interface $z = 0$ and develops into the bulk. (b) The integrated order parameter $|\Psi|(T)$ (Eq. S8) plotted against the temperature. The inset shows the high temperature region and signifies the first-order phase transition as well as the change from a concave to convex curvature.

Furthermore, we assume for the bulk a free energy including a quadratic, cubic and quartic term, and for the surface a quadratic free energy, as is done by Lipowsky et al., leading to:

$$f(\Psi, T) = \frac{1}{2}a\Psi^2 - \frac{1}{3}b\Psi^3 + c\Psi^4 \text{ and } f_s(\Psi, T) = \frac{1}{2}\alpha_1\Psi^2 \tag{S6}$$

The quadratic from of $f_s(\Psi, T)$ in Eq. S6 is the simplest assumption of surface free energy. To induce surface ordering, the phenomenological parameter $\alpha_1$ must be smaller than zero. Otherwise, the minimum of $f_s(\Psi, T)$ at the surface would be at $\Psi = 0$, which would lead to surface melting. Furthermore, we would like to draw attention to the temperature dependence of the free energy per area, $F(\Psi, T)$. Even though a surface term was introduced, the only temperature dependence is in the bulk term $a = a_o(T - T_0)$.

The order parameter profile $\Psi(z)$ can be derived by using the above two equations and taking $\Psi_s = \Psi(0)$ as a boundary condition. This results in:

$$\Psi(z) = \frac{2a}{\frac{2}{3}b + \sqrt{\bar{R}}\sinh\left(\sqrt{\frac{a}{J}}z + S\right)}, \tag{S7}$$

where $\bar{\bar{R}} = 2c(a - a^*)$ with $a^* = a(T_m) = \frac{2b^2}{9c}$ for the bulk and $S = \sinh^{-1}\left(\frac{1}{\sqrt{\bar{R}}}\left(\frac{2a}{\Psi_s} - \frac{2}{3}b\right)\right)$ with $\Psi_s = \Psi(0)$ being the order parameter at the surface $z = 0$. The order parameter profile, $\Psi(z)$, is plotted against z for varying temperatures in Fig. S18 (a). The dark red and blue curve have the highest and lowest temperatures, respectively. In our experiments, we cannot observe the order parameter profile directly, rather we observe the average order parameter across the entire film thickness. Therefore, to determine the temperature dependence of the average order parameter, we must integrate the order parameter profile $\Psi(z)$ over the entire film thickness. By integrating $\Psi(z)$, we obtain:

$$\Psi = \sqrt{\frac{2J}{c}}\ln\left(1 + \frac{\sqrt{\frac{a}{A^*}}}{\frac{1}{\sigma} + \sqrt{\frac{1}{\sigma^2} - \frac{a}{a^*}\left(1 - \frac{2}{\sigma}\right)} - 2\sqrt{\frac{a}{a^*}}}\right), \tag{S8}$$

*Contact author: oleksandr.dolynchuk@physik.uni-halle.de

where $\sigma = 1 + \sqrt{\frac{1}{A^*}\left(\frac{\alpha_1}{J} - a\right) + 1}$. Fig. S18(b) shows $\Psi$ plotted against temperature. It is clear that it is concave at higher temperatures, transitioning to convex when the order parameter begins to diverge into the bulk. However, our experimental data for P3HT and PDPP[T]$_2$-T indicate that $\Psi$ is convex at high temperatures. Furthermore, as can be seen in the inset, the theory predicts a weak first-order phase transition, whereas the experimentally obtained order parameter decreases continuously. Therefore, Lipowsky's theory cannot account for the behavior of our data and thus excludes the first-order phase transition in P3HT and PDPP[T]$_2$-T.

**Derivation of equilibrium order parameter $|\Psi|(T)$ by analogy to nematic LCs in an electric field**

We assume that the surface induces an edge-on predominant molecular orientation that creates a local molecular field, which then triggers the formation of positional order. This field breaks director symmetry and can be considered linear in $|\Psi|$. Similarly to the description of nematic LCs in the field [40,41], adding a negative linear term to the standard expression for free energy density that describes weak first-order phase transitions yields:

$$f(|\Psi|, T) = f_0 - h|\Psi| + \frac{a}{2}|\Psi|^2 - \frac{b}{3}|\Psi|^3 + \frac{c}{4}|\Psi|^4 \qquad \text{(S9)}$$

To avoid overparametrization, we introduce new parameters as follows: $\tilde{a} = \frac{a}{c} = \frac{a_0}{c}(T - T_0) = \tilde{a}_0(T - T_0)$, $\tilde{b} = \frac{b}{c}$, $\tilde{h} = \frac{h}{c}$. Minimizing the free energy with respect to $|\Psi|$ yields a cubic equation, which can be solved by reducing it to a depressed cubic with $|\Psi| = t + \frac{\tilde{b}}{3}$

$$0 = t^3 + pt + q$$

where $p = \tilde{a} - \frac{1}{3}\tilde{b}^2$ and $q = -\frac{2}{27}\tilde{b}^3 + \frac{1}{3}\tilde{a}\tilde{b} - \tilde{h}$. Finally using Vieta's substitution, $t = w - \frac{p}{3w}$, we get a quadratic equation and then solving it, the equilibrium order parameter is obtained as:

$$|\Psi|(T) = w - \frac{p}{3w} + \frac{\tilde{b}}{3}, \qquad \text{(S10)}$$

where $w = \left[-\frac{q}{2} + \left(\frac{q^2}{4} + \frac{p^3}{27}\right)^{\frac{1}{2}}\right]^{\frac{1}{3}}$.

| Sample | $\tilde{h}$ | $\tilde{a}_0$ | $\tilde{b}$ | $T_0$ | $T_m = T_0 + \frac{2\tilde{b}^2}{9\tilde{a}_0}$ | $\lvert\Psi\rvert(T_m) = \frac{2\tilde{b}}{3}$ | $\Delta T_{max} = \frac{\tilde{b}^2}{4\tilde{a}_0}$ |
|---|---|---|---|---|---|---|---|
| P3HT $L$=139nm | 0.014 | 0.036 | 0.379 | 246.665 | 247.552 | 0.253 | 0.998 |
| PDPP[T]$_2$-T $L$=286nm | 0.054 | 0.036 | 0.245 | 296.662 | 297.033 | 0.163 | 0.417 |
| PDPP[T]$_2$-T $L$=469nm | 0.098 | 0.022 | 0.754 | 285.893 | 291.636 | 0.503 | 6.460 |

**Table S2:** The fitting parameters $\tilde{h}$, $\tilde{a}_0$, $\tilde{b}$, and $T_0$ obtained for the red fitting curves in Fig. 4 are listed together with the calculated from them bulk melting temperature $T_m$, order parameter jump at the bulk melting temperature $|\Psi|(T_m)$, and the maximum bulk supercooling, $\Delta T_{max}$ (see page 16 above).

As can be seen in Table S2, the $\tilde{h}$ parameter increases four to seven times for PDPP[T]$_2$-T relative to P3HT. This indicates a stronger molecular field in PDPP[T]$_2$-T, as expected based on our experimental data. Furthermore, parameter $\tilde{b}$, which determines the jump of the calculated order parameter $|\Psi|(T_m)$, takes on reasonable values. For P3HT, the calculated value of the order parameter $|\Psi|(T_m)$ in Table S2 agrees quite well with its experimental estimate in Fig. S19(b). Since no isotropic LCs were observed during the cooling or heating of PDPP[T]$_2$-T films, no quantitative statement can be made about its $|\Psi|(T_m)$. The calculated bulk melting temperature $T_m$ of both polymers also shows comparable agreement with the experimental values. However, the $T_m$ of the thin PDPP[T]$_2$-T film is notably higher than $T_m$ of the thick film, which agrees well with the experimental estimate of about 289°C [33]. Additionally, the maximum supercooling values $\Delta T_{max}$, dependent on $\tilde{a}_0$ and $\tilde{b}$, differ by an order of magnitude from their experimental estimates for P3HT and thin PDPP[T]$_2$-T films but are again reasonable for thick PDPP[T]$_2$-T films. The discrepancy observed in both cases is likely due to the lower maximum fitted value of $|\Psi|^2$ in the region of lower temperatures in Fig. 4, which is critical for accurately determining $\tilde{a}_0$ and $T_0$. Finally, we should also bear in mind the impact of several factors on the experimental data in Fig. 4, including the finite accuracy of our measurements and the calibration value $I_{(100)}^{max}$ used to obtain $|\Psi|^2$, and other instrumental factors. One should also consider the possible limited applicability of the theoretical approach used when judging the accuracy of the fitting parameters in Table S2.

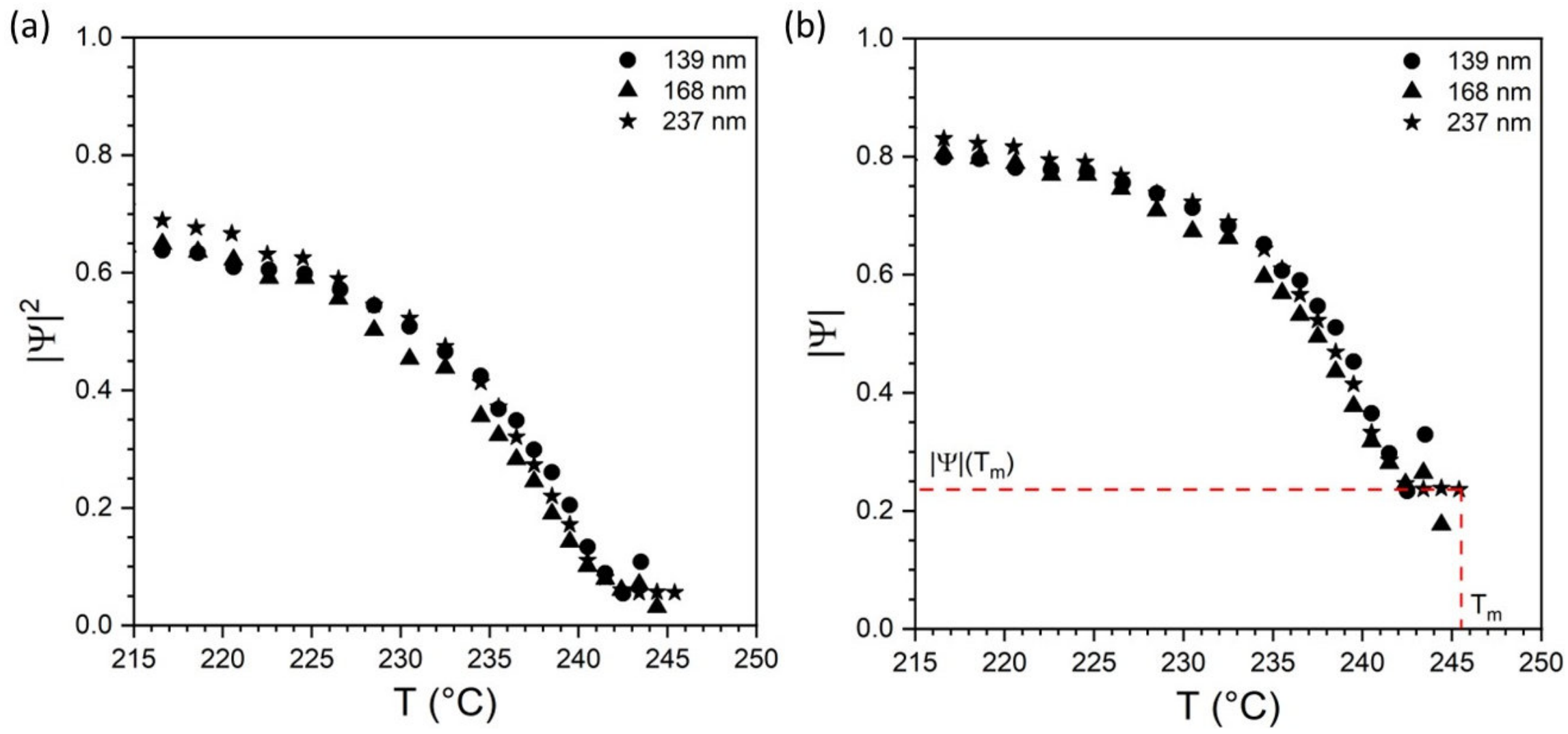

**Fig. S19:** (a) The squared smectic order parameter against temperature for the bulk in P3HT thin films (Figure 1 (e)) with the film thicknesses indicated in the legend. The squared smectic order parameter is calculated by $\frac{I_{(100)}(T)}{I_{(100)}^{max}} = \frac{N_{(100)}(T)\cdot|F_{(100)}(T)|^2}{N_{(100)}^{max}\cdot|F_{(100)}^{max}|^2} \approx |\Psi|^2$ (b) The smectic order parameter plotted against temperature for the same films. The dotted lines are guides to the eye to indicate the melting temperature and order parameter at the melting temperature.

*Contact author: oleksandr.dolynchuk@physik.uni-halle.de